\documentclass[%
aip,%
graphicx,%
amsmath%
]{revtex4-1}
\usepackage{color}
\usepackage{graphicx}
\usepackage{xr} 
\def\bea {\begin{eqnarray}}
\def\eea {\end{eqnarray}}
\def\be {\begin{equation}}
\def\ee {\end{equation}}

\begin{document}

\title{Emergence of Landauer Transport from Quantum Dynamics: A Model Hamiltonian Approach} 
%
%
%
\author{Partha Pratim Pal}
\author{S. Ramakrishna}
\author{Tamar Seideman}
\email[]{t-seideman@northwestern.edu}
\affiliation{Department of Chemistry, Northwestern University, Evanston, IL 60608, USA}
\maketitle 
%
%
\section{Introduction}
\label{Intro}
The interest in time dependent (TD) quantum transport has been growing with the ever improving technologies. Although nanoscale devices have long been fabricated and characterized, it is only recently that the challenge of combining the spatial  with temporal resolution to capture short time phenomena was achieved. 
Ultrafast experiments were initially restricted to solid state thin films but are now increasingly manifested in molecular scale materials. 
Irrespective of the system, the electrons are first to react to the ultrashort perturbations and may or may not attain a dynamic equilibrium depending on the time span of the disturbance. 
Theoretical methods for exploring the steady-state electron flow (dynamic equilibrium) in mesoscopic systems is a mature field and the challenge is gradually shifting towards molecular and atomic scale systems. 
Although several of the established mesoscale methods are transferable to the sub-nanometer scale, junction conductance is completely oblivious to it. 
Quantum tunneling, which dominates the transport characteristics in sub-nanoscale junctions, requires a different set of theoretical tools. 

The scenario becomes more complicated if the flow of electrons acquire a time dependence, for example, if they are driven by an alternating voltage or if they interact with an ultrashort laser pulse.  
In general, it is difficult to employ a steady-state formalism to describe transience\cite{Carey2017} but the reverse mapping can be natural and convenient.  
With this motivation, the theory of TD quantum transport has become an active area of research during the past few decades.  
Early efforts were directed towards developing a TD Green function framework that would converge seamlessly into a Landauer type conductance formula under certain conditions\cite{Jauho1994}. 
The advent of powerful computers lead to numerical recipes for computing TD Green's functions\cite{Hou2006, Moldoveanu2007, Prociuk2008, Gaury2014, Ridley2015, Ding2016, Popescu2016} in open systems. Computationally expensive time dependent density functional theory (TDDFT) was also used as an approach to calculate the TD current for both   open\cite{Wang2013a} and closed systems\cite{Stefanucci2004}.  
In the simplest implementation, the one-particle density matrix was propagated in time by a driving term in a tight-binding model with mean-field interactions at the level of adiabatic local density approximation\cite{Sanchez2006}. 
First-principles realization of a similar methodology was reported recently\cite{Morzan2017}. 
A quasi-steady-state current-voltage characteristics was obtained in a finite sized nano junction using TDDFT on a tight-binding model within a microcanonical approach which agreed with static scattering calculations\cite{DiVentra2004, Bushong2005}.
More recently, with a more generalized implementation of the above framework, Ercan and Anderson showed that the emergence of TD current and local occupation functions are structure dependent\cite{Ercan2010}. 
TDDFT based studies were reported, for example, to explore the transient current in molecular devices under ac bias\cite{Baer2004, Wu2005, Ke2010}, to calculate currents due to a pulsed\cite{Zhu2005} or an exponentially turned on voltage\cite{Wang2011a}, or an optical pulse\cite{Galperin2008}, and to illustrate dynamic suppression of current due to charge build up\cite{Evans2009}. 
The effects of the lead size\cite{Cheng2006} and the TDDFT exchange and correlation functionals on the transient current \cite{Evans2009a} were explored in refs.~\onlinecite{Cheng2006, Evans2009a}.
Open boundary conditions were implemented within TDDFT using a modified Crank-Nicholson algorithm and a stationary current was achieved for one dimensional model systems\cite{Kurth2005}. 
Using a microcanonical framework within TDDFT, similarities between pressure gradients in classical fluid flow and transient electronic current in nanoscale junctions were illustrated\cite{Sai2007}.
Comparison of transport characteristics as obtained from static DFT + NEGF and TDDFT using complex absorbing potentials to tackle open boundary conditions were performed in benzene-dithiolate and bipyridine junctions with gold electrodes\cite{Varga2011}.
A combination of TDDFT and NEGF within the wide-band limit approximation was also used to investigate transient transport characteristics in realistic molecular junctions\cite{Zheng2007, Zheng2010, Zhang2013b}.
Recently, a TDDFT approach based on grids in real space and real time was applied to a system of conjugated molecules and reliability of the concept was established\cite{Schaffhauser2016}.
A combination of quantum dissipative theory and TDDFT was also utilized to study optically induced transient current in a weakly coupled molecular device\cite{Cao2015}.
In another recent study, the dependence of early transient transport on the symmetry of the initial state of the system was explored\cite{Tu2016}.
Simulations using density functional based tight binding (DFTB), which are an order of magnitude cheaper than TDDFT, also appeared to capture the transient current dynamics although only at low biases\cite{Wang2011}.

Mapping the real-space to state-space representation and propagating the density matrix via the Louiville-von Neumann equation of motion can also capture the transient transport and, moreover, allow modeling of complex multi-terminal finite but realistic junctions\cite{Zelovich2014, Chen2014, Zelovich2016, Zelovich2017}.
Other formalisms include a many body path integral Monte Carlo approach that, when applied on a quantum system coupled to a phonon bath exhibited a TD current that reached a non-equilibrium stationary state after the initial transience\cite{Muhlbacher2008}.  
In another many body approach the Kadanoff-Baym equations for the NEGF were propagated in time resulting in the inclusion of electronic correlations and the ability to include TD external perturbations\cite{Myohanen2009}. This method not only estimated the transient current and molecule charging times\cite{Myohanen2009} but was also able to address the molecule-lead effect on the screening and the relaxation times\cite{Myohanen2012}.
The effect of electronic correlations (both electron-electron and electron-vibration) on the transient current was also explored through multilayer multiconfiguration time dependent Hartree theory\cite{Wang2013}.

Most of the transient current features reported in the literature can however be captured within a wave function approach (rather than density propagation), which (subject to certain approximations discussed below) is computationally cheaper. 
In addition, the interaction terms of the molecular bridge with the electrodes can be extracted from first-principles calculations (formally at any level of accuracy) and incorporated into the dynamics. 
In this article, we formulate and investigate quantum transport by applying the time dependent Schr\"{o}dinger equation to a model Hamiltonian.
Our Hamiltonian is an extended version of the one used earlier to study heterogeneous electron transfer\cite{Ramakrishna2001}.
We focus on simple two-terminal model devices, where the electrodes and the bridge are replaced by uniformly spaced quasi-continuum (QC) and a set of discrete energy levels respectively. 
Our model Hamiltonian consists of separate sets of energy levels corresponding to each sub-system (left and right lead, molecular bridge) and coupling terms between the molecule and the quasi-continua.
At the initial time, the populations in the quasi-continua are determined by the Fermi-Dirac distribution (FDD) with a common Fermi level ($E_f$).

We construct the most general expression for the TD current ($I_L$) in terms of the rate of change of population in the left QC (LQC), which has two components $I_{in}$ and $I_{out}$ corresponding to the incoming and the outgoing populations, respectively.  
Our calculations demonstrate (1) at early times $I_{in}$ and $I_{out}$ oscillate with a damped frequency and amplitude,
(2) after a certain time, both $I_{in}$ and $I_{out}$ settles into a constant  value and $I_L$ matches exactly with steady-state Landauer current. 
We term the commencement of the overlap between the two currents (exact and Landauer) as the onset of the Landauer regime and are the first ones to show the natural emergence of Landauer transport from a first-principles quantum dynamical picture sans prior assumptions. 
The initial frequency and amplitude of the oscillations and the timescale of the onset of the Landauer regime depend strongly on the electronic coupling strength of the molecule bridge with the electrodes.
We are able to define and extract a time dependent transmission (and reflectance) function and illustrate that it transitions into a steady state after the onset of the Landauer regime. Finally, we focus on the importance of our framework by comparing the current calculated by our exact expression to the Landauer current when time dependence in the electronic coupling between the bridge and the electrodes is turned on. 
Additionally, we examine the charge dynamics when hot-electrons are generated via plasmonic excitations in one of the electrodes.  

The paper is organized as follows: In section~\ref{formalism} we introduce the quantum dynamical model and the main working equations. Subsequently, we provide the results and discuss them extensively in section~\ref{RD}. Finally, in section~\ref{conclusions}, we conclude the paper with a brief summary and outlook of our work.

\section{Formalism}
\label{formalism}
\subsection{Transport from quantum dynamics}
\label{tqd}
A two-terminal device is modeled by three sub-systems - two leads, each replicated by a uniform QC and a bridge state (BS) described by a discrete set of energy levels. Thus, the Hamiltonian of the above system is written as:
\begin{equation}
\begin{split}
H =
\displaystyle\sum_{k_l}\epsilon_{k_l}\mid k_l~\rangle \langle~k_l \mid + 
\displaystyle\sum_{k_r}\epsilon_{k_r}\mid k_r~\rangle \langle~k_r \mid +
~\displaystyle\sum_m\epsilon_{m}\mid m~\rangle \langle~m \mid \\
+\bigg(\displaystyle\sum_{k_l} V_{k_l m}\mid k_l~\rangle \langle~m \mid +
\displaystyle\sum_{k_r} V_{k_r m}\mid k_r~\rangle \langle~m \mid +~h.c\bigg)
\end{split}
\label{hamiltonian}
\end{equation}
where $\epsilon_{k_{l/r}}$ and $\mid k_{l/r} \rangle$ are the energy eigenvalues and eigenstates respectively of the electrodes. 
$\epsilon_m$ and $\mid m \rangle$ are the stationary energy level and state of the BS. $V_{k_{l/r} m}$ are the terms between coupling the BS to the quasi-continua. The total wave function is written as:
\begin{equation}
\begin{split}
\mid \psi(t)_{total} \rangle = \displaystyle\sum_mc_m(t) e^{-i\omega_m t} \mid m \rangle + \displaystyle\sum_{k_l} c_{k_l} e^{-i\omega_{k_l} t}\mid k_l \rangle + \displaystyle\sum_{k_r} c_{k_r} e^{-i\omega_{k_r} t}\mid k_r \rangle,
\end{split}
\label{totalwavefunction}
\end{equation}
where $c_{k_{l/r}}$ and $c_m$ are the complex amplitudes corresponding to each eigenstate in the QCs and the BS respectively and $\omega_x=\epsilon_x/\hbar, x\in m,k_l,k_r $. Substituting equation~\ref{totalwavefunction} in the time dependent Schr\"{o}dinger equation and 
using the orthogonality of the $\mid x\rangle$ we obtain:
\begin{equation}
\begin{split}
i\hbar~\dot{c_{k_l}} = \displaystyle\sum_m c_m V_{k_l m}e^{i(\omega_{k_l} - ~\omega_m)t}
\end{split}
\label{lcoeff}
\end{equation}
\begin{equation}
\begin{split}
i\hbar~\dot{c_{k_r}} =  \displaystyle\sum_m c_m V_{k_r m}e^{i(\omega_{k_r} - ~\omega_m)t}
\end{split}
\label{rcoeff}
\end{equation}
\begin{equation}
\begin{split}
i\hbar~\dot{c_m} =  \displaystyle\sum_{k_l}V_{k_l m}c_{k_l} e^{i(\omega_{k_m} - \omega_{k_l})t} + \displaystyle\sum_{k_r}V_{k_r m}c_{k_r} e^{i(\omega_{k_m} - ~\omega_{k_r})t} .
\end{split}
\label{mcoeff}
\end{equation}
Again, from the Heisenberg equation of motion we derive the following equation for the expectation value of the rate of change of occupancy in the  LQC:
\begin{equation}
\begin{split}
\langle\dot{n}_L^{k^0_{l/r}}\rangle~=~-\dfrac{i}{\hbar}\big[n_L^{k^0_{l/r}}, H\big]
~= ~\dfrac{2}{\hbar}Im\big\{\displaystyle\sum_{k_l}V_{k_l m}c_{k_l}^*c_me^{i(\omega_{k_m} - ~\omega_{k_l})t}\big\},
\end{split}
\label{he}
\end{equation}
where $\langle n_L^{k^0_{l}}(t)\rangle = \displaystyle\sum_{k_l}|c_{k_l}(t)|^2$ with the initial condition - ONE eigenstate is populated @t=0 and the expectation value is obtained from that TD wave function. Formally, equation~\ref{he} can be rewritten in a condensed form as: 
\begin{equation}
\begin{split}
\langle\dot{n}_L^{k^0_{l/r}}\rangle~ =~\dfrac{2}{\hbar}Im\big\{\displaystyle\sum_{k_{l/r}}T_{k_{l/r}m}\big\}f_{k^0_{l/r}}~=~S_{k^0_{l/r}}f_{k^0_{l/r}},
\end{split}
\label{he1}
\end{equation}
where $f_{k^0_{l/r}}$ is the Fermi-Dirac occupancy (@t~=~0) of a particular eigenstate in either QC. 
We justify the factorization of equation~\ref{he1} as a product of Fermi-Dirac occupancy and a reflectance/transmission function in the appendix. 
If the initially occupied state is in the LQC (RQC), $S_{k^0_l}$ ($S_{k^0_r}$) is a reflectance (transmission) function. 
As the initial state of the full system is an incoherent superposition of the electrode states, the observables are determined by time propagating the wave function for a range of initial electrode states and summing the initial-state-dependent-observables with weights determined from a FDD. 
The electronic current in the LQC, denoted as $I_L$, can now be split into two contributions, one corresponding to initial population in the LQC ($I_{out}$) and the other for the population originating in the RQC ($I_{in}$):
\begin{equation}
\begin{split}
I_{out}~=~\sum_{k_l^0}\langle\dfrac{d}{dt}n_L^{k^0_l}\rangle~=~\sum_{k_l^0}S_{k^0_l}f_{k^0_l}, \\
I_{in}~=~\sum_{k_r^0}\langle\dfrac{d}{dt}n_L^{k^0_r} \rangle~=~\sum_{k_r^0}S_{k^0_r}f_{k^0_r},
\end{split}
\label{he2}
\end{equation}
where $\sum_{k_{l}^0}$ and $\sum_{k_{r}^0}$ represents the summation over all initial conditions associated with the initial populations at LQC and RQC respectively. 
The total current in the left electrode $I_L$ can then be written as:
\begin{equation}
\begin{split}
I_L=I_{out}-I_{in}=\sum_{k_l^0}S_{k^0_l}f_{k^0_l}-\sum_{k_r^0}S_{k^0_r}f_{k^0_r}.
\end{split}
\label{he3}
\end{equation}
which is our most general expression for the TD current valid at {\it all} times. 

\section{Results and Discussion}
\label{RD}
\subsection{Dynamics}
\label{dynamics}

We investigate the population dynamics within our single-particle framework by introducing a finite number of particles into our microcanonical (total energy and the number of particles is fixed) system.
At zero bias, the equilibrium Fermi levels ($E_f$) in both quasi-continua are identical and are placed at the center of the range of QC levels (0 - 1.2 eV  with a spacing of $\Delta_{QC}= 0.005$~eV unless or otherwise mentioned) to avoid band edge effects and allow comparison with the analytical data. 
We restrict our time propagation to $t < t_r (=\frac{2\pi\hbar}{\Delta_{QC}})$, where $t_r$ is the recurrence time - defined as the time limit until which a QC acts as a realistic continuum\cite{Ramakrishna2001}.
The maximum level spacing we use in our calculations is $\Delta_{QC}= 0.005$ and so we cutoff all our quantum dynamics simulation at 800 femtoseconds (fs) or even before. 
The population ($\langle{n}_x^{k^o_{l/r}}\rangle; x\in$ BS, LQC, RQC) as a function of time in the various sub-systems for energy independent coupling strengths ($V_{k_{l/r}m}$) and BS on-resonance with $E_f$, is illustrated in Figure Supplementary Information (SI)-2. 
%
A certain amount of population is transferred from the QCs to the empty BS until the populations in the sub-systems achieve a dynamic equilibrium. 
The overall population decay rate of the two QCs are identical since there is no external bias on the system and $V_{k_{l}m}=V_{k_{r}m}$. 
The saturation population at the BS and the equilibration time are determined by $V_{k_{l/r}m}$, the density of states ($\rho$) in the quasi-continua ($\rho={\Delta_{QC}}^{-1}$) and the alignment of the BS with $E_f$. 
In the example shown here the saturation population is 0.5 and the equilibration time is $\sim 391$~fs. 
The saturation population on the BS is directly proportional to $V_{k_{l/r}m}$ and is maximum if the BS is on-resonance with $E_f$. 

\begin{figure}[h!]
 \includegraphics[width=0.75\textwidth]{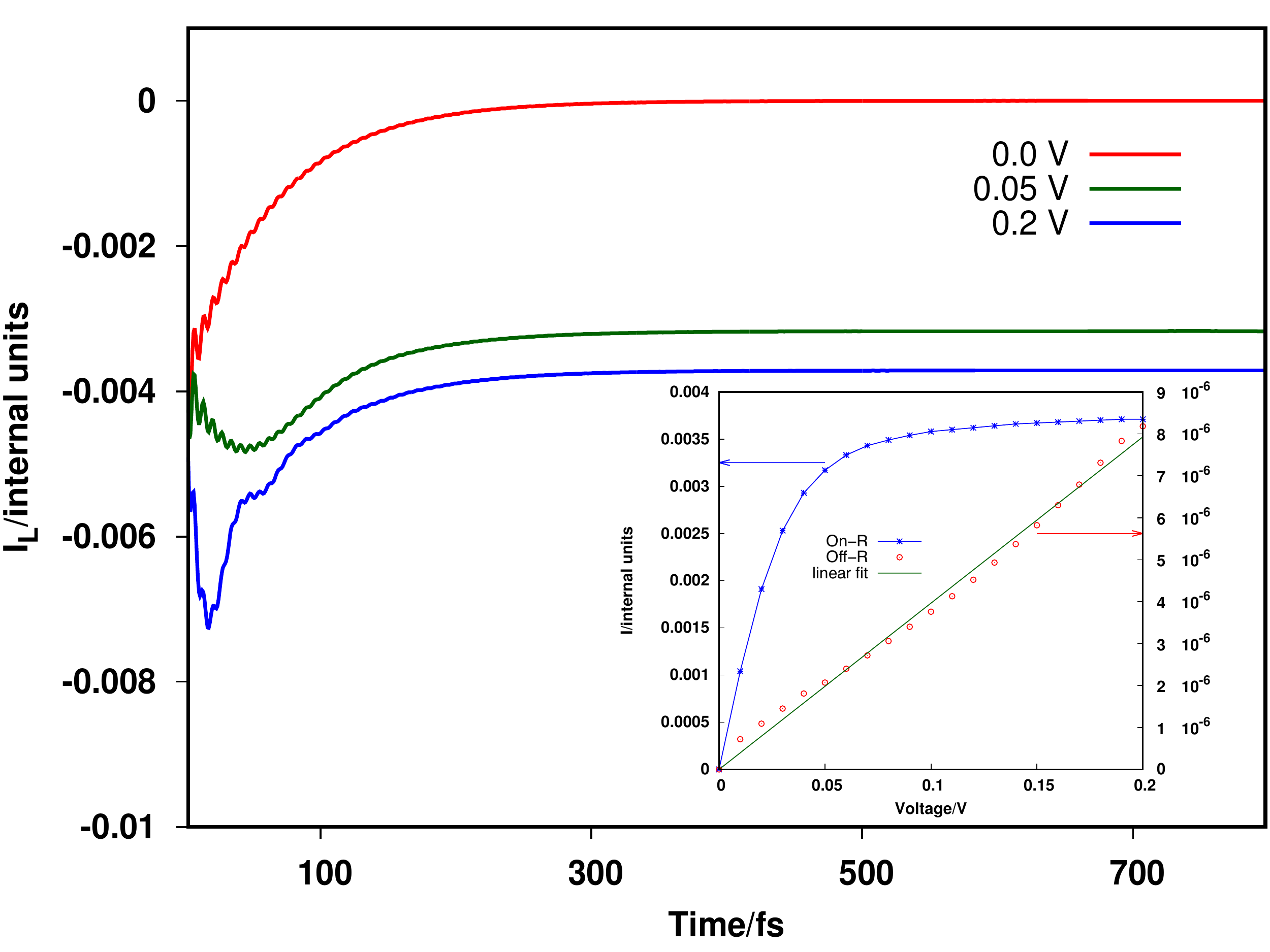}
 \caption{$I_L$ as a function of time for different external voltages as mentioned in the legends. Inset: current-voltage characteristics when BS is on (on-R) and off-resonance (off-R) with $E_f$. QC levels run from $0 - 1.2$ eV with a spacing of 0.005 eV with equilibrium Fermi level at 0.6 eV. BS is positioned at 0.6 eV and 1.0 eV for the on-R and off-R cases respectively. $V_{k_{l/r} m}$~= 0.002 eV for both the arrangements.  
}
 \label{ILtotal}
\end{figure}

We introduce an external bias ($V_B$) by modifying the local chemical potential (CP) in the QCs by $\pm V_B/2$, where the positive sign corresponds to a higher CP at the left ($\mu_1=E_f+V_B/2$) compared to that in the right electrode ($\mu_2=E_f-V_B/2$).  
In the presence of bias voltage, $I_L$  (defined in equation~\ref{he3}) 
saturates, after short time fluctuations, to a non-zero as seen in Figure~\ref{ILtotal}. As expected $I_L$ settles to 0 (red) at $V_B$ = 0 V. 
$I_L$ is negative at long times because $V_B$ is positive resulting in steady-state unidirectional population flow from the LQC to the right QC (RQC).    
Similar quasi-steady currents in closed systems have been reported earlier in several TD quantum transport calculations on model junctions\cite{Zhu2005, Kurth2005, Bushong2005, Sanchez2006, Zheng2007, Varga2011, Yam2011, Wang2011a, Schaffhauser2016}. 
The long time limit of $I_L$ is the non-equilibrium steady-state current and  it is evidently bias dependent as shown in the inset of Fig. \ref{ILtotal}.
Near resonance (blue line in the inset of Figure~\ref{ILtotal}) we observe a sigmoidal current-voltage characteristics. The same voltage dependence was found in early steady-state calculations of the transport\cite{DiVentraM2000, Pal2011} and of current-driven dynamics\cite{Alavi2000}. The trend is readily explainable through the analytical theory of \cite{Alavi2000} to mirror the eigenvalue structure of the bridging molecule, with the steepness of the sigmoidal curve determined by the resonance lifetime.  
If the BS is outside the window ($E_f + 0.4$ eV) of $\mu_1$ and $\mu_2$ (off-R) the current-voltage alliance is linear (red dots with a green fit), which is consistent with off-resonance tunneling trends\cite{Pal2010}.
Thus our formulation, along with our definition of current approves the steady-state current-voltage relationship reported in the literature. 
It is however interesting to note that the increase in the steady-state current with $V_B$ is maximum in the on-R model, where BS $=(\mu_1+\mu_2)/2$. 
Shifting the BS (denoted by $\Delta$) diminishes the slope of the sigmoidal current-voltage curve as shown in Figure SI-3 
although the long-time limit current at higher applied biases converges to the same value for all BS locations.  

\begin{figure}[h!]
\includegraphics[width=0.75\textwidth]{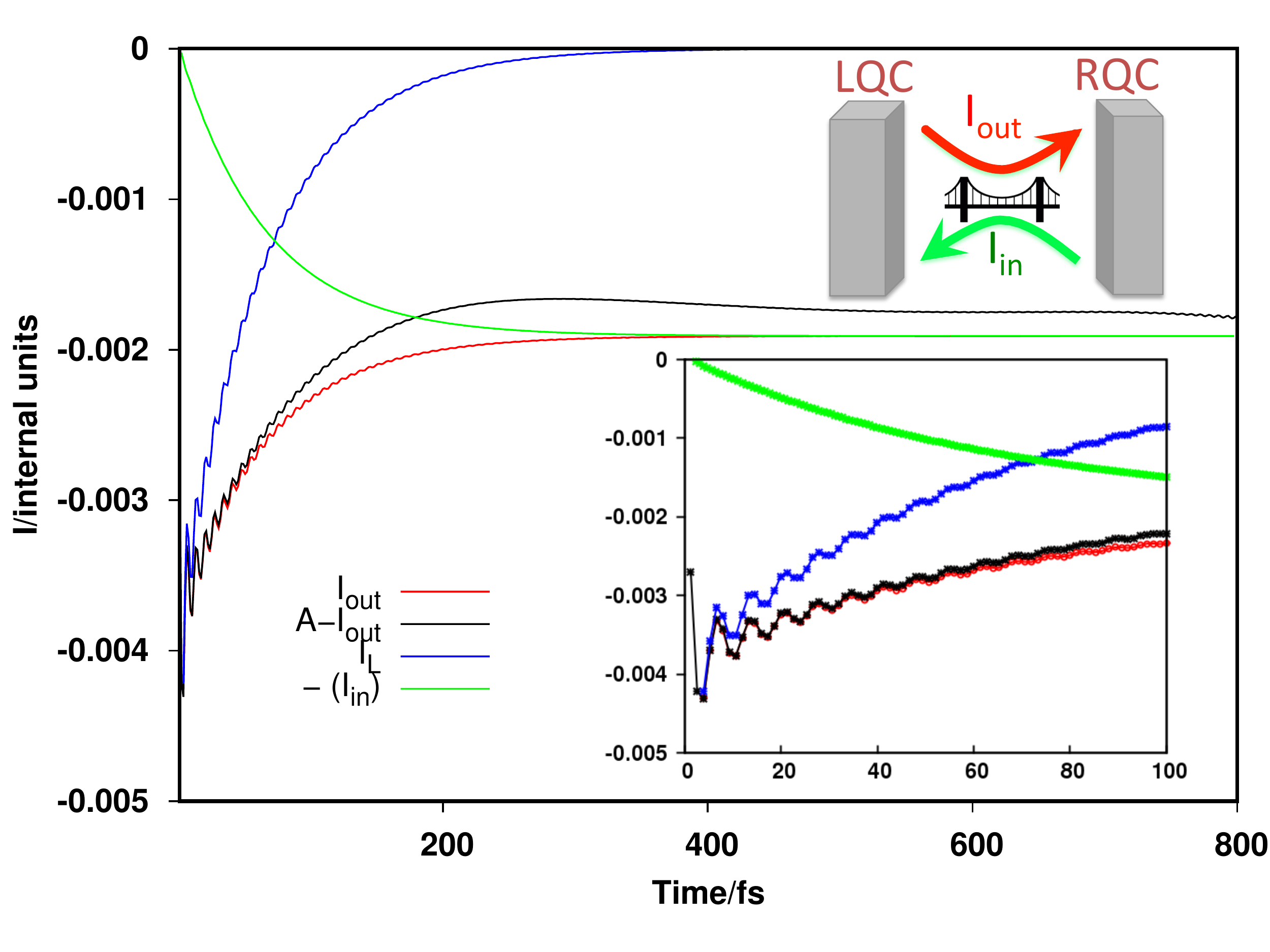}
\caption{Current dynamics at the LQC and its $I_{out}$ and -$I_{in}$ components. Bridge level (BS) is on-resonance with the equilibrium $E_f$ of the QCs. The definition of the components are provided in the text. Inset: Transient current at very early times. The parameters are identical to the on-R model in Figure~\ref{ILtotal}.
}
\label{mp-on}
\end{figure}

\begin{figure}[h!]
\includegraphics[width=0.75\textwidth]{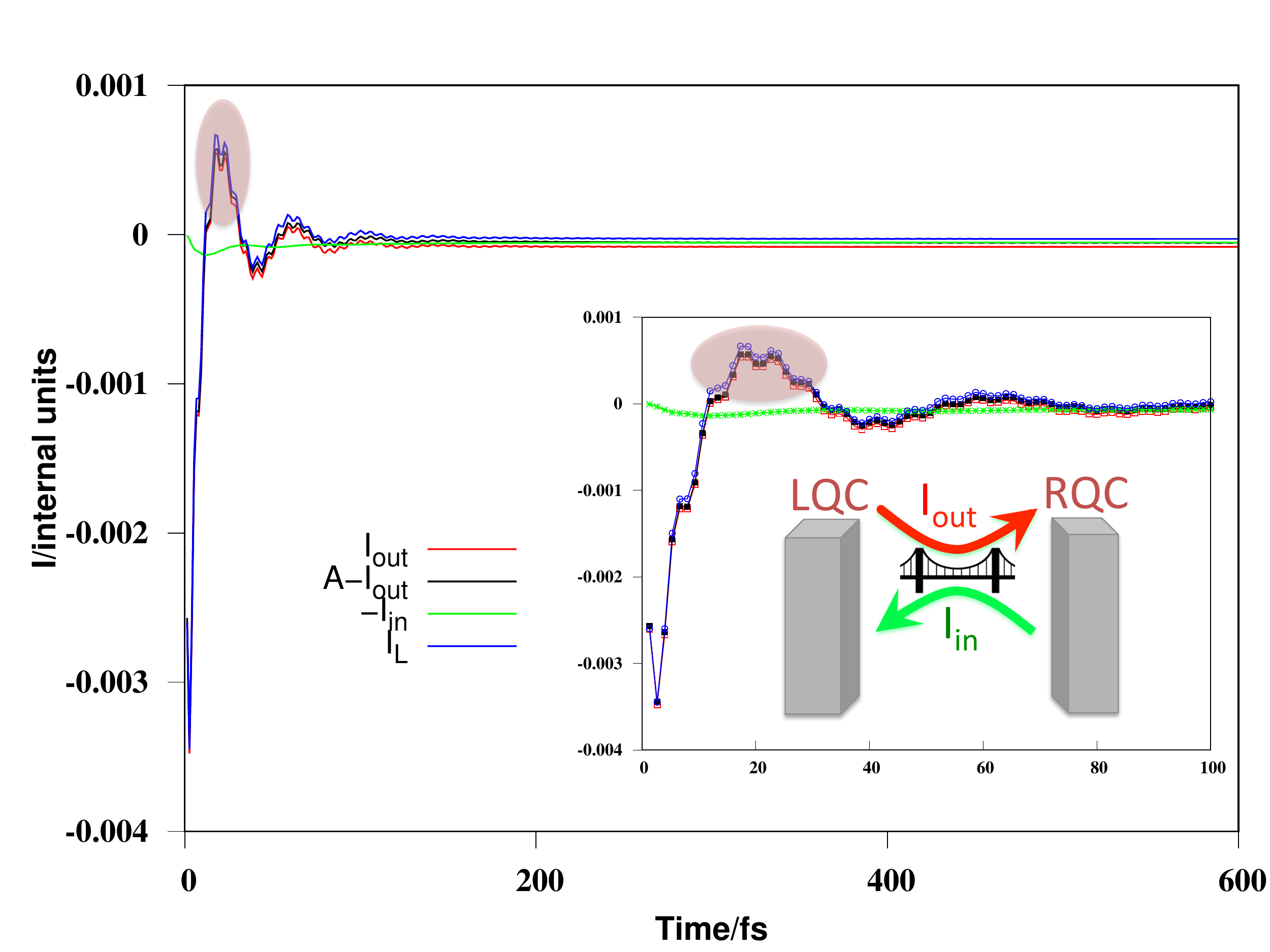}
\caption{Dynamical current at the LQC and its $I_{out}$ and -$I_{in}$ contributions. Inset: Current at very early times. Parameters are identical to Figure~\ref{mp-on} but BS is located at 0.7 eV.
}
\label{mp-poff}
\end{figure}

\subsection{Non-equilibrium Steady State}
I$_L$ consists of 2 components, one ($I_{out}$) corresponding to initial population in the left and the other ($I_{in}$) to the initial population in the right QC. 
In Figure~\ref{mp-on}, we show the $I_L$ and its components at $0~V$ for the on-R model introduced in section~\ref{dynamics}. 
$I_{out}$ is always negative because the population flux originating in the LQC leaks out of it at all times. In contrast, $I_{in}$ is always positive since it adds population to the LQC. However, we plot $-(I_{in})$ to illustrate that $I_{in}$ and $I_{out}$ settle into an equal and opposite value and at the same time, which is 391 fs for this particular arrangement of energy levels. 
Beyond this point, the transient features in the current (and its components) disappears and the rate of population exchange attains a steady value (for $V_B\neq0$V,) replicating the time-independent non-equilibrium steady-state current.  
Thus, commencing from a quantum dynamical description of population transport between the LQC-BS-RQC we arrive at a time independent charge flow, which is a measurable quantity.  

It is however interesting to note that $I_{out}$ oscillates at early times (inset of Figure~\ref{mp-on}) due to significant backscattering from the BS to the LQC, whereas $I_{in}$ exhibits a relatively smooth trend.  
This behavior is consistent with an earlier observation that an electron when traveling from a narrow conductor to a wide lead suffers from minimal reflection\cite{Szafer1989} but the vice-versa is not true\cite{Datta1995, DiVentra2008}.  
The segment of population originating in the RQC and leaking out of LQC is negligible because coupling between the sub-systems is relatively weak. Oscillations in $I_{in}$ however observed if the coupling strength is significantly increased (for e. g., a 5 fold increase) as illustrated in the results of Figure SI-4.

\subsection{Comparison with an Analytical Theory}
We derive an analytical expression for $I_{out}$ to arrive at a structure for a more generalized expression for the time-dependent electron transport, which is later shown to reduce to the Landauer expression in the steady state. We are interested to explore the physics of the analytical expression, though they are admittedly approximate, as the truncated recursions ignore the population dynamics of the initially unoccupied states in the dynamical expression for the initially occupied states. This neglect propagates even through the dynamics of the initially unoccupied states that are derived subsequently from the incomplete expression for the initially occupied states as elaborated in the appendix.

The black line (denoted as A-$I_{out}$) in Figure~\ref{mp-on} corresponds to $I_{out}$ calculated from the analytical expression (derived in the appendix) for the on-R arrangement. Our analytical theory agrees well with the numerical calculations (red) for the initial few tens of fs (inset of Figure~\ref{mp-on}) after which the high-frequency current fluctuations evaluated from either expression are damped and additionally, they start to deviate from each other. 
This discrepancy arises because we stop at the first recursion to get to a closed form of the analytical expression as detailed in the appendix. A consequence of this approach is the exclusion of at least two higher order processes: (1) influence of the initially unoccupied LQC levels on the rate of population leakage at the initially occupied level in the same electrode, (2) the population dynamics of a specific initially unoccupied LQC level being exclusively related to the population of the initially occupied level and not on the rest of the unoccupied LQC levels.  
Both the above-mentioned higher order processes, which are unaccounted for in our simplified analytical expression, involve backscattering of the BS population into the LQC levels. The time scale of backscattering  associated with a single LQC level can be approximated by $2\pi\hbar/\Delta_{BS-\epsilon_{k_l}}$, where $\Delta_{BS-\epsilon_{k_l}}$ is defined as the energy difference between BS and $\epsilon_{k_l}$.
However, it is difficult to nail the time scale of the overall backscattering that disturbs the initial overlap between the numerical and the analytical currents because of the summation over all $\epsilon_{k_l}$ in $I_{out}$(equation~\ref{he2}). 

In the off-resonance case (Fig.~\ref{mp-poff}), where BS $= E_f + 0.1$~eV, the analytical expression for I$_{out}$ (black) is in excellent agreement with its numerical counterpart (red). 
This result could be anticipated - the backscattering of BS population into LQC in this situation is inefficient because of lack of BS population in the first place (note - BS is on-resonance with initially unoccupied LQC levels). 
As a result, the higher order processes ignored by our restricting the analytical solution to the first recursion, are unable to play a dominant role in the overall population dynamics. 
We also note (see Figure~\ref{mp-poff}) that $I_{L}$ is strongly dominated by I$_{out}$ in the off-resonance case.   
Due to lack of occupied RQC levels around BS, I$_{in}$ is negligibly small at early time and converges to zero as time progresses.    
It is also interesting to note that $I_L$ in this setup reverses its direction (as marked by shaded oval areas in Figure~\ref{mp-poff}) before the onset of steady state.
To analyze this behavior we replot the dynamical current alongside the rate of change of population in the LQC level resonant with BS (marked by a magenta line `X' in the schematic of Figure SI-5).
Part of the BS population is backscattered into `X', culminating in LQC  population gain until $\sim$150 fs before the rate of gain starts to drop off. 
We establish this by tracking $I_{out}$ in Figure SI-5
, which changes its sign synchronously with $I_L$ around 20 fs. 
In the inset of the same figure, we further observe the on-resonant LQC level gains population on a similar time scale, thus supporting the sign reversal of $I_{out}$.
When the bridge level is below the Fermi level (BS $= E_f - 0.4$ eV) the current dynamics (Figure SI-6)
is similar to Figure~\ref{mp-on}. The initial oscillations in $I_L$ are, however, less frequent and more damped, possibly due to a sizable amount of population exchange between the initially filled levels of both the quasi-continua and the BS. 
The analytical expression in this scenario agrees with the numerical $I_{out}$ at early time scales and displays low frequency oscillation instead of transitioning into a steady state. 
Overall, we do establish that our truncated analytical expression is able to capture the transience although does not perform well in the long-time limit. However, the structure of our analytical formula cements the foundation for the rest of the interrogations in the article. 

\begin{figure}[h!]
 \includegraphics[width=0.75\textwidth]{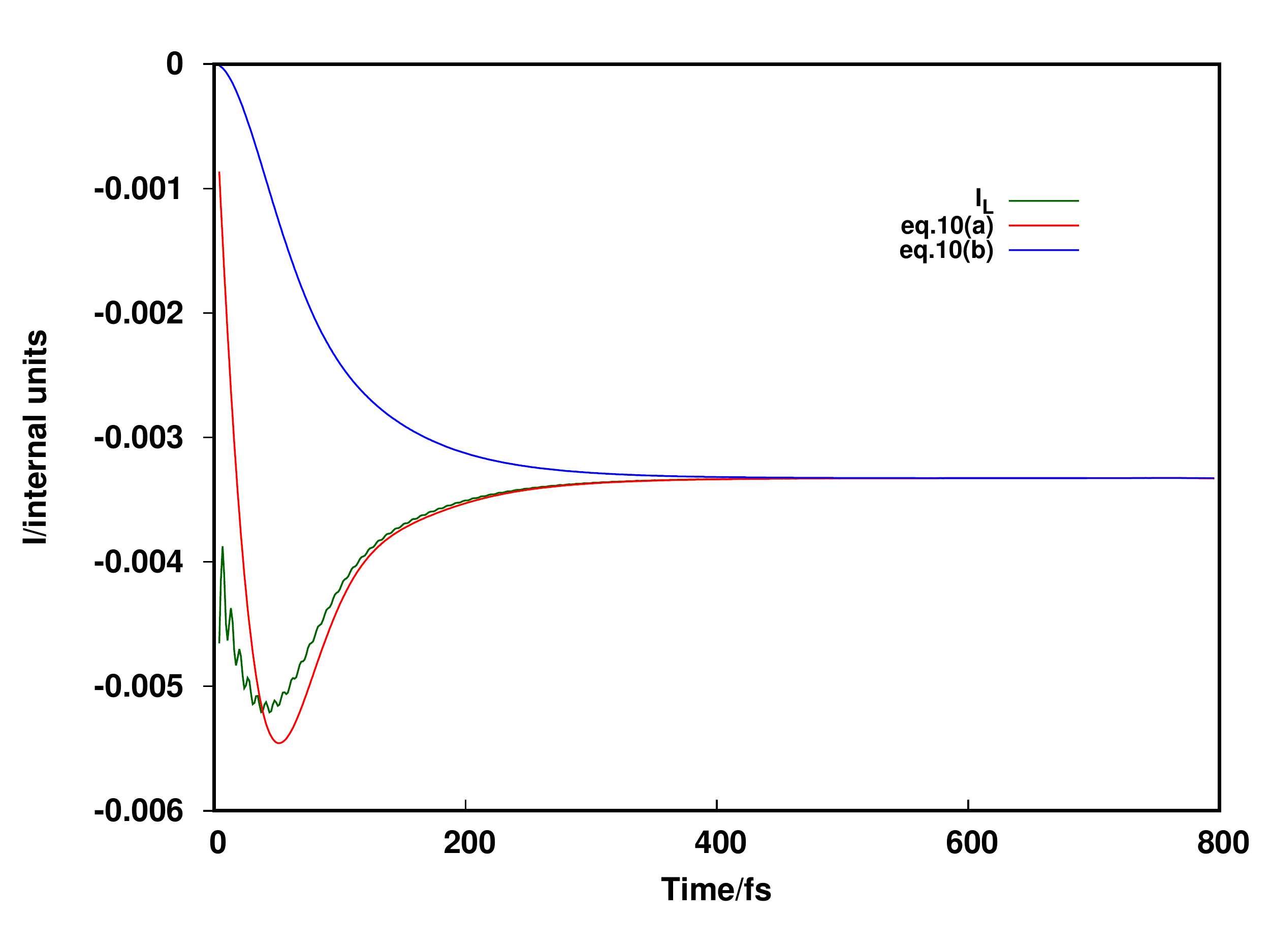}
 \caption{$I_{in}^{Land}$ and $I_{out}^{Land}$ as calculated from an approximate Landauer type equation~\ref{he4}. The exact $I_L$ is also plotted to illustrate the emergence of the Landauer regime. The model and parameters of Figure~\ref{ILtotal} are used but with $V_B$ = + 0.06 V}
 \label{approx_L}
\end{figure}

\begin{figure}[h!]
 \includegraphics[width=0.75\textwidth]{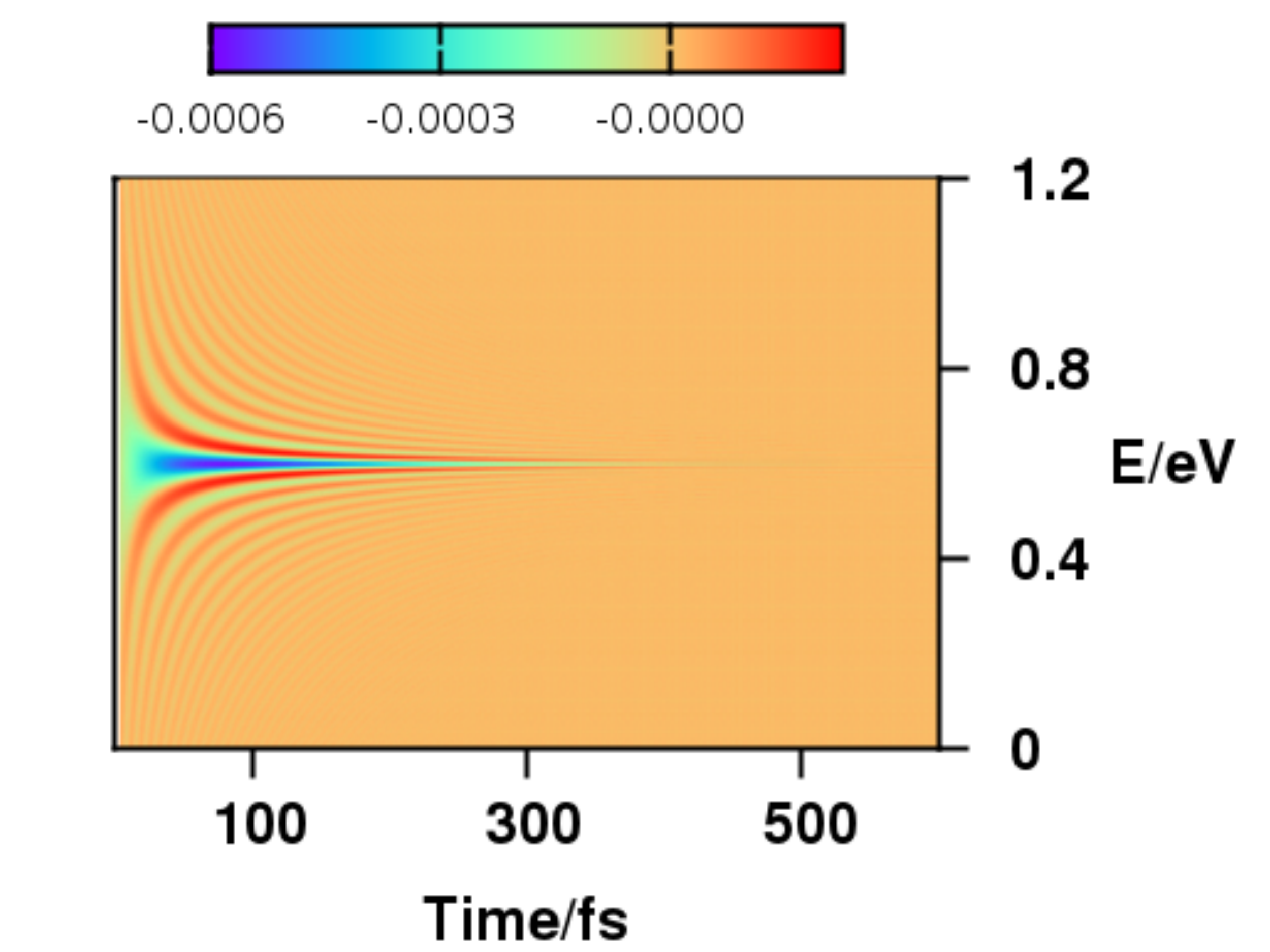}
 \caption{Difference of the time dependent functions $S_{k^0_l}$($\epsilon$,t) and $S_{k^0_r}$($\epsilon$,t). The model and parameters of Figure~\ref{ILtotal} are used here.
 }
 \label{diff_transm}
\end{figure}

\begin{figure}[h!]
 \includegraphics[width=0.75\textwidth]{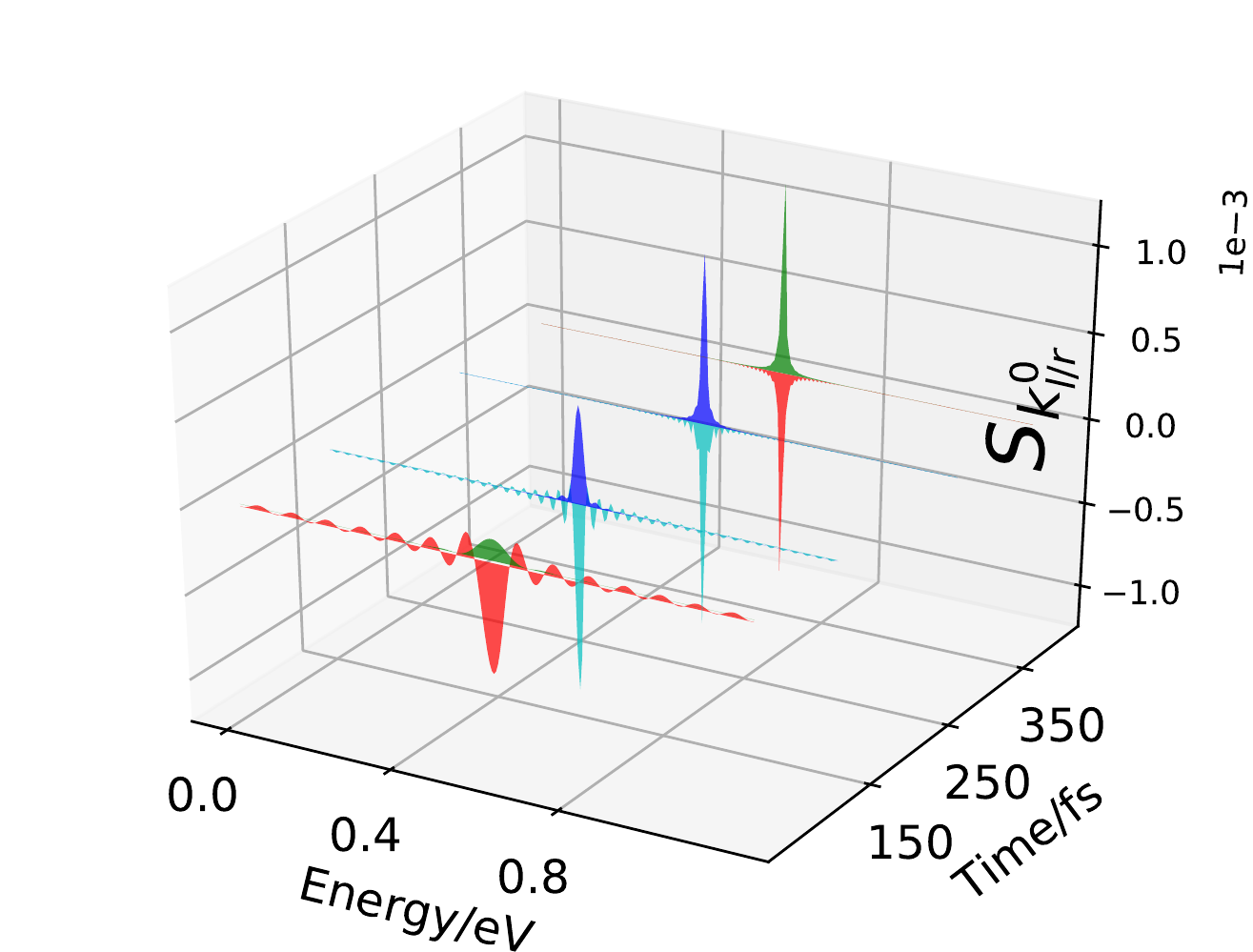}
 \caption{Time dependent reflectance(red and cyan) and transmission(blue and green) functions, $S_{k^0_{l/r}}$($\epsilon$,t). The model and parameters of Figure~\ref{ILtotal} are used here. 
 }
 \label{transm}
\end{figure}

\begin{figure}[h!]
 \includegraphics[width=0.75\textwidth]{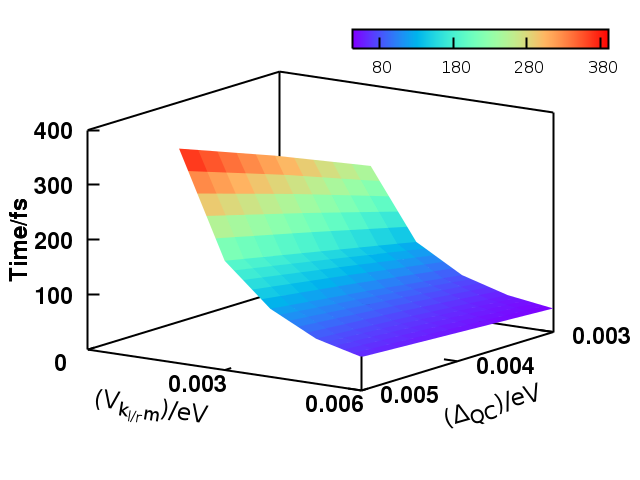}
 \caption{Onset of Landauer regime for a range of symmetric coupling  strength between the bridge and the QC and energy levels spacing in the QCs. The bridge level is on-resonance with the E$_{F}$.}
 \label{vc-multi}
\end{figure}

\begin{figure}[h!]
 \includegraphics[width=0.75\textwidth]{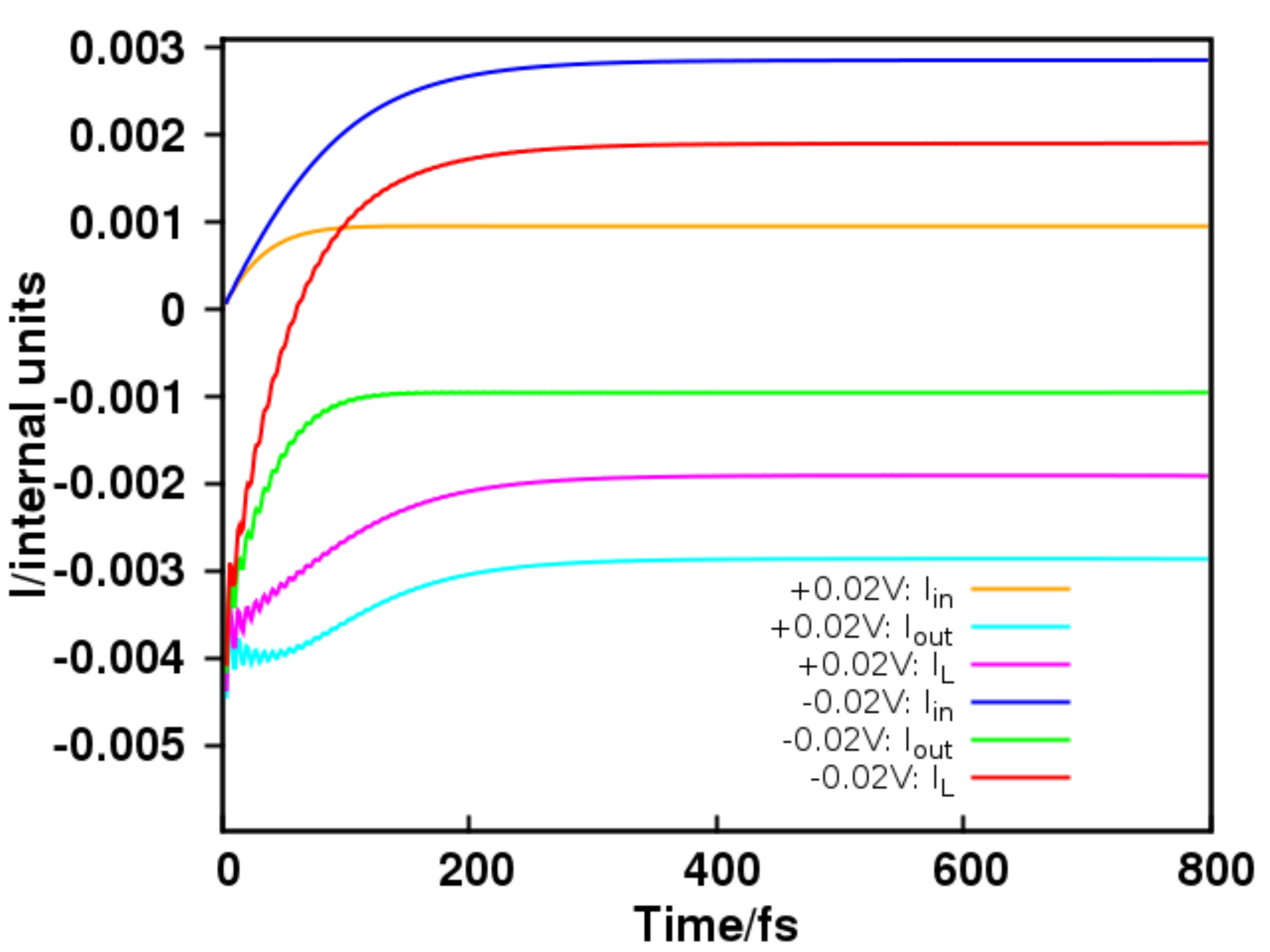}
 \caption{Onset of Landauer regime for a range of symmetric coupling (C) strength between the bridge and the QC and energy levels spacing (S) in the QCs. The bridge level is on-resonance with $E_f$.}
 \label{Iin}
\end{figure}

\subsection{Onset of the Landauer Regime}
The analytical solution of $I_{out}$ in equation~\ref{SklFkl} (Appendix) indicates that the FDD can be factored out as a time independent segment of the function. $I_{in}$ is also expressible similarly (equation~\ref{SkrFkr} in Appendix) and exploiting this structural form we approximate I$_{L}$ as:
\begin{subequations}
\begin{eqnarray}
I_L=\displaystyle\sum_{k_l^0}S_{k^0_l}\big(f_{k^0_l}-f_{k^0_r}\big)=I^{Land}_{out} \label{he4a} \\
=\displaystyle\sum_{k_r^0}S_{k^0_r}\big(f_{k^0_r}-f_{k^0_l}\big)=I^{Land}_{in}
\label{he4b}
\end{eqnarray}
\label{he4}
\end{subequations}
The equations above are very similar to the Landauer formula:
\begin{equation}
\begin{split}
I_L=\displaystyle\sum_{\epsilon=-\infty}^{+\infty}T(\epsilon)\bigg[f_l(\epsilon)-f_r(\epsilon)\bigg],
\end{split}
\label{le}
\end{equation}
where $f_{l/r}(\epsilon)$ is the FDD at the left/right electrode. 
Comparing equations~\ref{he4} and~\ref{le}, we have that $S_{k^0_{l/r}}$ takes the role of a reflectance/transmission function in our formalism. 
In Figure~\ref{approx_L} we plot $I_{out}^{Land}$ and $I_{in}^{Land}$ for the on-R model, which are approximate Landauer currents given by equation~\ref{he4}. 
Red, blue and dark-green curves corresponds to $I_{out}^{Land}$, $I_{in}^{Land}$, and $I_L$ (equation~\ref{he3}) respectively. 
All the calculations are performed with $V_B = + 0.06$ V as equations~\ref{he4a} and ~\ref{he4b} go to zero at all times when $V_B = 0$ V.  
We clearly observe that the approximate Landauer type expressions are devoid of any rapid oscillations and are unable to capture the transient current.
Despite this drawback, the current via this expression converges to the correct $I_L$ rendering the Landauer type equations invalid prior to the onset of the steady state at 391 fs. 
Thus, we are able to capture the natural emergence of the Landauer transport from a quantum dynamical framework without the prior assumption of time independent charge flow between the sub-systems.  
This is the central result of our efforts here, which was possible only because of our analytical expressions that allowed us to define the Landauer $I_L$ and compare it with the numerically exact $I_L$.

To analyze the emergence of the Landauer regime, we plot the difference of the transmission ($S_{k^0_r}$) and the reflectance function ($S_{k^0_l}$) as a function of time in Figure~\ref{diff_transm}. 
At an early time the difference as a function of energy is finite but this quantity goes to zero after the onset of the Landauer regime with $S_{k^0_{l}}=S_{k^0_{r}}$ at all energies, further illustrated in Figure~\ref{transm}. 
$S_{k^0_{r}}$ is plotted in green and blue in Figure~\ref{transm} whereas red and cyan curves show $S_{k^0_{l}}$. 
$S_{k^0_{l}}$ values are negative because it corresponds to population decrease in the LQC.  
Both transmission and reflectance functions at early times are spread over multiple energy levels before settling down to equal and opposite Lorentzian functions (last data point) centered around the BS after the population exchange attains a dynamic equilibrium. 
The full-width at half maxima (fwhm) of the Lorentzian depends on the electronic coupling between the QC and the BS. 
After the onset, one of the $S$'s in equation~\ref{he3} can be factored out resulting in the Landauer type formulas of equation~\ref{he4}.  
Increasing $V_{k_l m}$ results in a steady-state $S_{k^0_r}$ (and also $S_{k^0_l}$)  with a larger fwhm as shown in Figure SI-7a
and an early Landauer onset (Figure SI-7b)
as detailed in the next section.

\begin{figure}[h!]
 \includegraphics[width=0.75\textwidth]{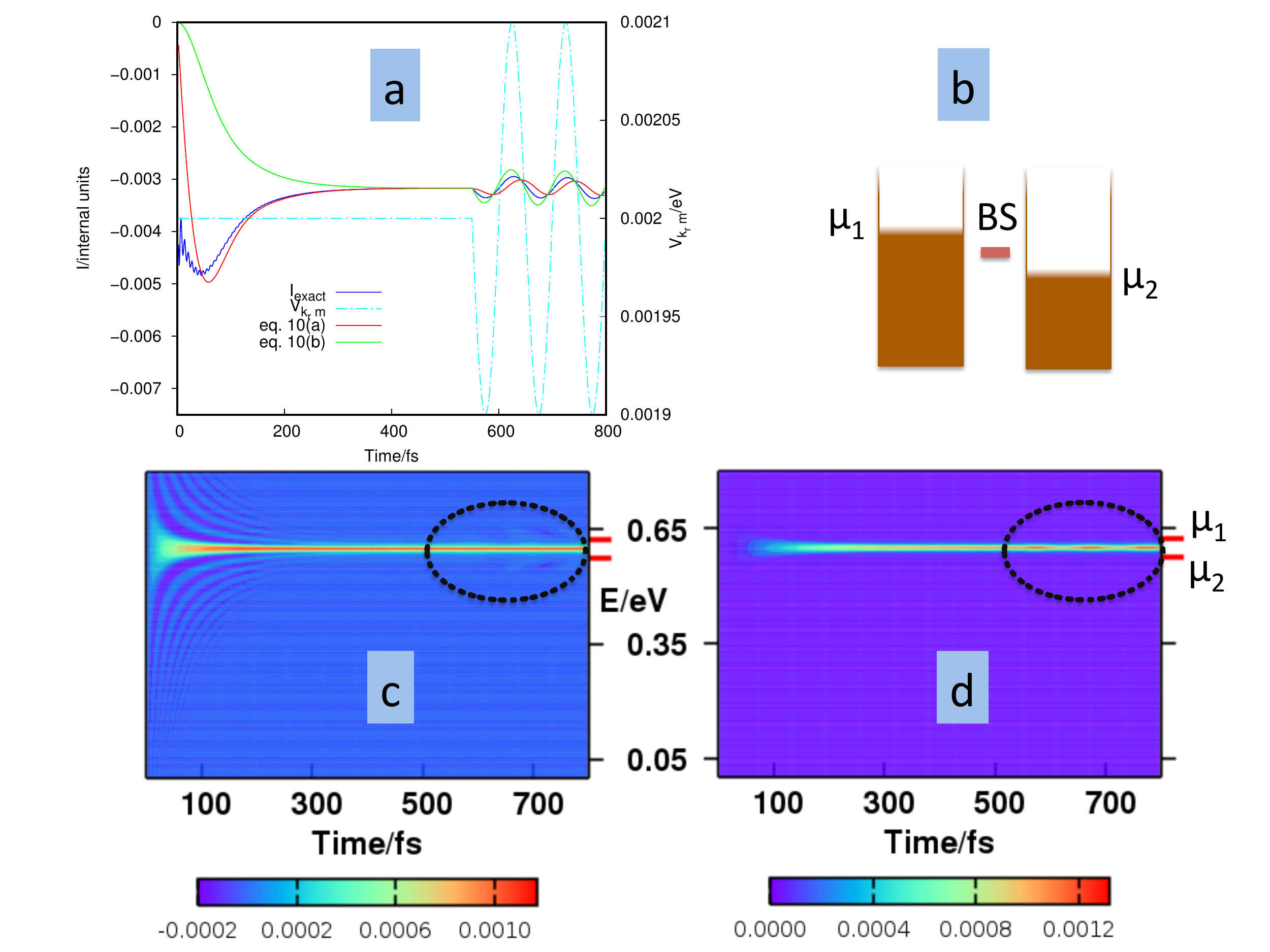}
 \caption{(a)Comparison of exact $I_L$ with approximate Landauer type equation~\ref{he4} in the presence of time dependent coupling ($V_{k_{l/r} m}$ (cyan curve plotted against right ordinate) between the bridge and the electrodes. (b) The position of BS and the chemical potentials are shown using schematics for the two respective situations. Bias = $\mu_1-\mu_2$ = 0.05 V. (c) Reflectance as a function of time and energy. (d) Transmittance as a function of time and energy. Dotted ovals in (c) and (d) indicate the region where the coupling is time dependent. $\mu_1$ and $\mu_2$ are marked by short red dashes in (c) and (d).
 }
 \label{td_coupling}
\end{figure}

\begin{figure}[h!]
 \includegraphics[width=0.75\textwidth]{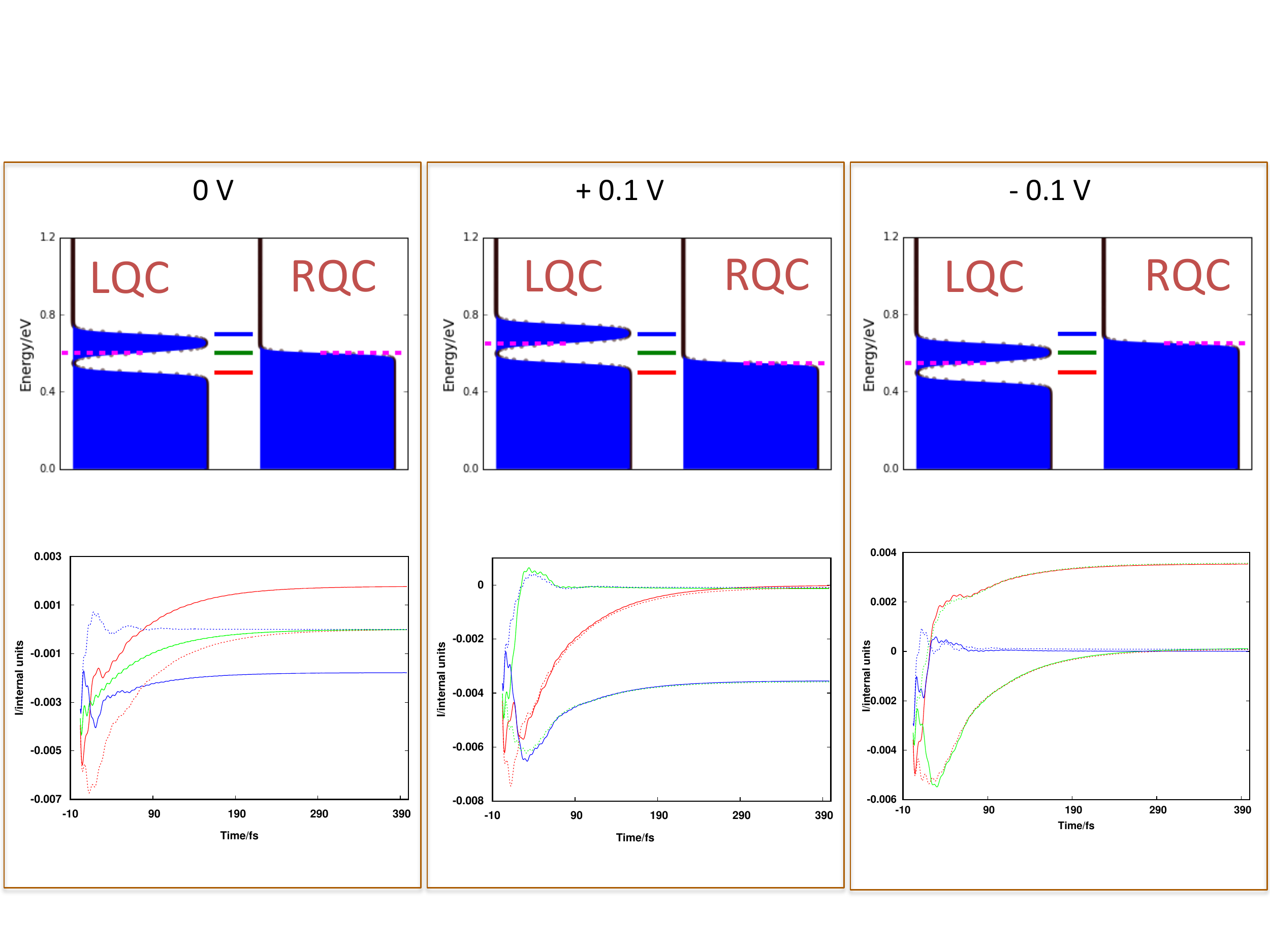}
 \caption{I$_L$ in the absence (dotted lines) and presence (solid lines) of hot electrons that are generated via plasmon decoherence in the LQC. Blue, green and red lines in the top panel represent the position of BS and the same colors are correspondingly used to plot the dynamical current in the bottom panel.}
 \label{hotE}
\end{figure}

\subsection{Controlling the Onset Time}
We now proceed to systematically calculate the onset time of the Landauer regime as a function of the coupling strength and the energy level spacing of the QCs for the on-resonant case and present it in the form of a 3D plot in Figure~\ref{vc-multi}. 
We define the onset of Landauer regime,  $t_L$, as the time when $I_L$ in equation~\ref{he3} goes to zero when $V_B= 0$ V.  
For a symmetric coupling of BS with the QCs, $t_L$ is inversely proportional to $V_{k_{l/r} m}$; for example, $t_L$ is $\sim$~391 and 82 fs for coupling strengths of 0.002 and 0.005 eV respectively. 
In the weak coupling regime, $t_L$ decreases as the energy levels in the QC are more closely spaced but as $V_{k_{l/r} m}$ gets stronger $t_L$ becomes independent of the level spacing.
Between the two, $t_L$ has a stronger dependence on $V_{k_{l/r} m}$ than $\Delta_{QC}$ because the rate of population exchange between the sub-systems is controlled predominantly by the coupling strength.
Computing the dependence of $t_L$ on $V_{k_{l/r} m}$ and QC level spacing using the approximate Landauer expression of equation~\ref{he4} with a small bias of $+0.02$ V (Figure SI-8)
reveals that the trend is qualitatively similar to Figure~\ref{vc-multi}. 

Despite being a purely theoretical aspect - we find it interesting that the individual contributions $I_{in}$ and $I_{out}$ attain the non-equilibrium steady state at the same time, only at $V_B=0$ V (Figure~\ref{mp-on}). 
When biased, the QC which has larger initial population in the energy levels close to BS, takes more time to attain the steady state. 
For example, when $V_B= +0.02$ V, $I_{out}$ (cyan curve in Figure~\ref{Iin}) takes a much longer time to settle into a steady state whereas $I_{in}$ (orange curve in Figure~\ref{Iin}) flattens out much earlier. 
The trend is reversed at $V_B= -0.02$ V where $I_{out}$ (green) transitions into the dynamic equilibrium much earlier that $I_{in}$(blue). 
However, $t_L$, which marks the onset of $I_L$ steady-state is independent of $V_B$.   
By contrast to the results of Figure~\ref{approx_L}, the individual contributions $I^{Land}_{in}$ and $I^{Land}_{out}$ in equation~\ref{he4} transition into the steady state at the same time ($=t_L$) even if $V_B\neq$ 0 V. 
In experiments, the measurements will detect $I_L$ only but $t_L$ may continuously vary in situations such as in a scanning tunneling microscope (STM)-break junction approach, where the interaction strength between the bridge and the electrode changes along the course of the experiment. 

\subsection{Time Dependent Coupling}
We next investigate the effectiveness of the Landauer formula in the presence of a TD perturbation\cite{Peskin2017} for example, when a vibrational wave packet is excited in the bridging molecule, its distance from either electrode varies sinusoidally as a function of time\cite{Kaun2005} and so also would the electronic coupling matrix elements. 
We introduce this time dependence into our model by adding a term to the coupling (perturbation to $V_{k_l m}$ and $V_{k_r m}$ are out of phase by $\pi$) after the system settles into the Landauer regime. The red and green curves in Figure~\ref{td_coupling}(a), which are calculated using equations~\ref{he4a} ($I_{Land}^{out}$)~and~~\ref{he4b} ($I_{Land}^{in}$) respectively, overlap with $I_{L}$ (blue curve) after $t_L$ and before the perturbation is applied. 
In Figure~\ref{td_coupling}(b) we provide a schematic used for the calculations, which is identical to the on-R model mentioned in section~\ref{dynamics}.
As soon as the time dependent term in $V_{k_{l/r} m}$ is switched on, $I_{in}^{Land}$ and  $I_{out}^{Land}$ drifts away from $I_{L}$ both in terms of amplitude and phase. $I_{out}^{Land}$ underestimates the current with a lower peak value as compared to $I_{L}$ whereas $I_{in}^{Land}$ overestimates. $I_{out}^{Land}$ is ahead in phase compared to $I_L$ and $I_{in}^{Land}$ is behind it. The disagreement between the analytical and numerically exact currents indicate the incompleteness of the Landauer expression when the coupling is time dependent. 

To account for the disparities between the two Landauer currents of  Figure~\ref{td_coupling}(a) we analyze the response of $S_{k^0_r}$ and $S_{k^0_l}$ to the sinusoidal component of the $V_{k_{l/r} m}$. After the time dependent term in the coupling is switched on, $S_{k^0_r}$ fluctuates about BS (located at 0.6 eV in Figure~\ref{td_coupling}(d)), the region indicated by dotted oval lines. In contrast, $S_{k^0_l}$ values at $\epsilon_{k_l}$ far off from BS also gains a time dependence (Figure~\ref{td_coupling}(c)). It is essential to note that equations~\ref{he4} and \ref{le} reduce to:   
\begin{equation}
I_L=\displaystyle\sum_{\epsilon_{k_l}=\mu_1}^{\epsilon_{k_l}=\mu_2}T(\epsilon_{k_l})\bigg[f_l(\epsilon_{k_l})-f_r(\epsilon_{k_r})\bigg].
\label{muL}
\end{equation}
at low temperatures, when the Fermi-Dirac distribution approaches a step function and thus, the terms inside the parenthesis go to zero for $\mu_2<\epsilon_{k_l}<\mu_1$.
As a result equation~\ref{he4a} captures the portion of the TD fluctuations in $S_{k^0_l}$ only within the range $\mu_1-\mu_2$ and misses the rest. However, equation~\ref{he4b} is able to capture the entire time dependence of $S_{k^0_r}$, which exclusively lies within the range $\mu_1-\mu_2$. This explains the existence of a higher peak in $I_{in}^{Land}$ and a smaller one in $I_{out}^{Land}$, with $I_L$ peaking  between the two $I^{Land}$ currents. 
Thus, our analysis elucidates the central reason behind the deficiency of the Landauer expression in this situation - the starkly different response of the reflectance and the transmission function to the time dependent coupling.  
 
\subsection{Hot-electron Dynamics}
In a scanning tunneling microscope, hot-electrons are created through plasmon decoherence when a laser is incident on a plasmonic tip. When the STM is engaged to a sample with monolayer of molecules hot electrons tunnel if they find suitable pathways such as molecular orbitals, giving rise to an additional current \cite{Pal2015}. 
In order to model this process using our time dependent framework, we assume the existence of hot electrons in the LQC. We borrow an analytical expression for the distribution from our previous work\cite{Kornbluth2013, Pal2015} to investigate the transient hot-electron dynamics.
The electrons in the RQC follow a pure FDD.
We calculate $I_L$ from equation~\ref{he3} for different BS positions, green for on-resonance and red (blue) for 0.1 eV below (above) the equilibrium Fermi level.
Using the same colors in Figure~\ref{hotE}, we plot $I_L$ as a function of time in the presence (absence) of hot electrons with solid (dotted) lines. 
The solid blue lines in the bottom panel of Figure~\ref{hotE}(a) indicate that when BS is 0.1 eV above $E_f$, current flows from LQC to RQC whereas when it is 0.1 eV below $E_f$, the electronic population flows from RQC to the empty levels in the LQC in the steady state (red solid).  
Thus, even at 0 V hot electrons generate a finite current, the direction of which depends on the position of BS. 
As expected, in the absence of hot electrons, $I_L$ (represented by dotted lines) goes to zero at steady-state in Figure~\ref{hotE}(a) irrespective of BS location. 
In the same figure, when BS is on-resonance with the equilibrium Fermi level, $I_L$ in the presence and absence of hot electrons (solid and dotted green) overlap with each other. 
This occurs possibly due to the proximity of BS to the tail of either the FDD or the hot electron distribution. 

At a finite bias ($+0.1$ V), the current may be suppressed to zero when plasmons (and subsequently hot electrons) are excited in the LQC, for example, red and green solid lines in Figure~\ref{hotE}(b). 
This happens because in both cases LQC and RQC levels close to BS are populated to similar extent; almost filled in red and almost empty for green. 
Similar considerations explain the results shown as blue and green solid lines in Figure~\ref{hotE}(c). In the same figure, the population flows from the right to the left QC (red solid) giving rise to positive $I_L$ when BS is below the Fermi level. This happens because LQC has vacant levels close to BS where the population from RQC can flow into.  
In the presence of bias (Figure~\ref{hotE}(b) and (c)), as expected from earlier sections, significant current flows in the absence of hot electrons only when BS lies within the window of $\mu_1$ and $\mu_2$. 
For all the cases considered here, $I_L$ transitions into the non-equilibrium steady state regime within few hundreds of fs and we verified that it converges to the Landauer current thereafter. This reinforces the validity of estimating the hot electron current via the Landauer expression in a STM tip-sample junction when the tip is excited by a continuous wave (CW) laser passed through a 80 Hz optic chopper\cite{Pal2015}. 
The CW laser is thus focussed on the STM tip for a few milliseconds during which the dynamical hot electron current would surely transition into the Landauer regime. 

\section{Conclusions}
\label{conclusions}
To summarize, we have devised a first-principles quantum dynamical framework for time dependent charge transport in a representative electrode-molecule-electrode arrangement. A quasi-continuum of evenly spaced energy levels and a single level depict the electrodes and the molecule respectively. 
Our approach is based on propagation of the single particle time dependent Schr\"{o}dinger wave function of a model Hamiltonian.   
Our calculations reveal a transient charge flow at early time scales prior to the onset of a steady state. Upon biasing, the steady-state  
current (defined as the rate of charge flow from the left electrode) attains a nonzero constant value and exhibits a qualitatively similar current-voltage characteristics as that obtained via the popular Landauer formula.  
To gain further insight into these results, we derive a formally exact analytical formula for the portion of the current due to population originating  from the left electrode, which shows excellent agreement with the numerical data at early time scales.
We construct a closed form approximation by truncating higher order processes in our exact analytical form, and illustrate the emergence of high order processes at long time, manifested as deviation of the closed form approximation from the numerical results. 
The structural form of the analytical expression reveals that the outgoing current can be written as the initial occupancy multiplied by a transmission or reflectance function.
Exploiting this observation, we rewrite the time dependent current in terms of an approximate Landauer style expression and illustrate that it matches  the exact current (obtained from the most generalized definition) after the population dynamics settles into a steady state.
We term this the onset of the Landauer regime, where, analogous to the Landauer formula, the net current in the left electrode can be expressed as a sum over the transmission (or reflectance) function weighted by the respective difference in the Fermi function occupancies.
In-depth analysis points out that the transmission and reflectance functions  transition into a time independent equal and opposite quantities after the onset.  
The onset time is however dependent on the coupling strength between the BS and the QCs and the QC level spacing with stronger coupling leading to faster onset of the Landauer regime. This behavior is explained in terms of rapid population exchange between the various subsystems.  

When the coupling between the BS and the QCs is time-dependent, for example, when a vibrational wave packet is triggered in the bridge of a molecular scale junction the numerically exact dynamical current deviates from the Landauer currents.  
Discrepancies are recorded both in amplitude and phase of the current and 
the disagreement is associated with the distinctly dissimilar response of the transmission and the reflectance function to the time dependence of the coupling. 
In the transmission case, the time dependent fluctuations occur only at energy levels close to BS whereas in the reflectance case the time varying coupling seems to have an effect at energy levels away from the BS. 
Through our formalism we are able to visualize hot-electron dynamics, which is again a time dependent phenomenon. 
We predict flow of electrical current in an unbiased junction triggered solely by hot-electrons that are created from plasmon decoherences in metal nanoconstructs (such as a STM tip). Similarly, the current in a biased junction can be suppressed by exciting hot-electrons in one of the electrodes. 
In both scenarios, we show the importance of the presence of an appropriate bridging level that mediates the population exchange between the electrodes. We further show that the hot-electron dynamics settles into a steady state within few hundreds of femtoseconds, after which the Landauer expression gives realistic values for the current. Thus, we find that the current generated via tunneling hot-electrons (created from laser excitations) can be estimated by the conventional Landauer formula. One of our goals in future research in this area would be to calculate the current under similar conditions but for realistic molecular junctions by extracting the parameters of the model Hamiltonian from first-principles simulations. Our work will thus have important implications for fundamental investigations in molecular optoelectronics. 

\begin{acknowledgments}
The authors thank Professors Mark Ratner and George Schatz for helpful discussions and the National Science Foundation (Grant No. CHE-1465201 and Grant No. DMR-1720139) for support. The numerical research reported used computational resources and staff contributions provided for the Quest high performance computing facility at Northwestern University, which is jointly supported by the Office of the Provost, the Office for Research, and Northwestern University Information Technology. Use of the Center for Nanoscale Materials, an Office of Science user facility, was supported by the U. S. Department of Energy, Office of Science, Office of Basic Energy Sciences, under Contract No. DE-AC02-06CH11357.
\end{acknowledgments}

\section*{Appendix: Analytical expressions for the microscopic currents}

We start from the differential equations for the TD amplitudes of $C_{k_l}(t)$, $C_{k_r}(t)$ and $C_m(t)$ namely equations (\ref{lcoeff}~-~\ref{mcoeff}) in the text.
Substitution, integration and by application of the initial condition $C_{k_l}(t=0)=\sqrt{f_{k_l}}$, yields [see equation~\ref{lcoeff}]
\bea
\label{full-ID}
\dot{C}_{k_l}(t)&=&-\frac{{|v|}^2}{\hbar^2}\left(\int_0^t dt'\,exp[-(A+iB)(t-t')]\,C_{k_l}(t')\right.\nonumber \\
&&+\left.\sum_{k_l'\neq k_l} e^{i\omega_{k_l}t}\int_0^t dt'\,exp[-(A+i\omega_m)(t-t')-i\omega_{k_l'}t']C_{k_l'}(t')\right). 
\eea
where $A=\frac{\pi v^2\rho}{\hbar}$, $B=(\omega_m-\omega_{k_l})$ and $\rho$ is the density of states in the right electrode. The density of states is 
inversely related to the quasicontinuum spacing i.e., $\rho={\Delta_{QC}}^{-1}$, where $\Delta_{QC}$ is the energy spacing between quasicontinuum levels. 
In arriving at the above we have made use of the Poisson summation~\cite{Lighthill1970}, $\sum_{k_r} e^{-i\omega_{k_r}(t-t')}=2\pi\hbar\delta(t-t')$, which is valid for an  
infinitely wide, uniform quasicontinuum of electrode levels, or the wide band limit (WBL). As the 
integro-differential equation obtained cannot be solved analytically, we attempt a recursive solution by ignoring levels that are initially unoccupied 
thus obtaining a closed form solution for the single level in the left electrode that is initially occupied. In other words, we assume that $C_{k_l'}(t)\approx 0$ in Equation~(\ref{full-ID})  
obtaining   
\be
\label{simp-ID}
\dot{C}_{k_l}(t)\approx -\frac{{|v|}^2}{\hbar^2}\int_0^t dt'\,exp[-(A+iB)(t-t')]\,C_{k_l}(t').
\ee
Equation (\ref{simp-ID})  is expected to be valid only in the short time limit.  
At long times neglecting the initially unoccupied levels will 
be incorrect as the probability amplitude in the initially occupied level will scatter back from the bridge level into the initially unoccupied left electrode levels.
It should be noted that when a term has a prime superscript, it basically represents the initially unoccupied level which enters the equations via its 
energy $\omega_{k_l'}$.
 The solution to Eq.~(\ref{simp-ID}) is obtained by Laplace transforms which results in 
\be
\label{occ-sol1}
C_{k_l}(t) \approx C^{(1)}_{k_l}(t)=\sqrt{f_{k_l}}(Le^{r_1 t} + Me^{r_2 t}),
\ee
and
\bea
L &=& \frac{(A+iB) +r_1}{r_1-r_2}, \\
M &=& \frac{(A+iB)+r_2}{r_2-r_1}, \\
r_1 &= & \frac{-(A+iB)+\sqrt{(A+iB)^2-4\kappa}}{2}, \\
r_2 &=& \frac{-(A+iB)-\sqrt{(A+iB)^2-4\kappa}}{2},
\eea
where $\kappa=\frac{v^2}{\hbar^2}$ and $C^{(1)}_{k_l}(t)$ denotes the first recursive solution pertaining to the assumption $C_{k_l'}(t=0)=0$.
Equation~(\ref{occ-sol1}) correctly 
reduces to $C_{k_l}(t=0)=\sqrt{f_{k_l}}$ at t=0 as one can verify that $L+M=1$.  As anticipated, as time progresses the
correct limit is no longer for two reasons; (1) as mentioned above, at long times probability amplitude is scattered back into unoccupied levels of the left electrode, a process that is not accounted for and (2) unphysical recurrences become non-negligible $t\geq \tau_r$, 
where $\tau_r=2\pi\hbar\rho$ is the recurrence 
time - a consequence of representing the electrode levels as a uniform QC [2]. 
The temporal behavior of the decay of the occupied level is controlled by 
 by three parameters: the electronic coupling strength ($V_{k_{l/r}m}$) between the electrode levels and the bridge, the density of states of the right electrode ($\rho$) and the energy difference between the electrode level and the bridge. The last parameter induces oscillatory behavior in the decay. 

In order to obtain the TD amplitudes of the initially unoccupied levels in the LQC at the first recursive , 
we can rewrite Equation~(\ref{full-ID}) as
\bea
\label{un-ID}
\dot{C}^{(1)}_{k_l'}(t)&= &-\frac{{|v|}^2}{\hbar^2}\left(\int_0^t dt'\,exp[-(A+iB')(t-t')]\,C_{k_l'}(t')\right.\nonumber \\
&&+\left. e^{i\omega_{k_l'}t}\int_0^t dt'\,exp[-(A+i\omega_m)(t-t')-i\omega_{k_l}t']C_{k_l}(t')\right),
\eea
where $C^{(1)}_{k_l'}$ represents the first recursive solution to an initially unoccupied level, while $C_{k_l}(t')$ in the 
above represents the amplitude of the initially occupied level. Again the dynamics of the other unoccupied levels have been 
neglected in Equation~(\ref{un-ID}). Substituting the first recursive solution for the occupied level, namely Eq.~(\ref{occ-sol1}), 
in Equation~(\ref{un-ID}), we obtain,
\bea
\label{un-sol1}
 C^{(1)}_{k_l'}(t) &=& \frac{P_1 L'}{r_{1}'-K_1}(e^{r_{1}'t}-e^{K_1 t}) + \frac{P_1 M'}{r_{2}'-K_1}(e^{r_{2}'t}-e^{K_1 t})\nonumber \\
&&-\frac{(P_1 + P_2)L'}{r_{1}'-K_2}(e^{r_{1}'t}-e^{K_2 t}) - \frac{(P_1 +P_2) M'}{r_{2}'-K_2}(e^{r_{2}'t}-e^{K_2 t})\nonumber \\
&&+\frac{P_2 L'}{r_{1}'-K_3}(e^{r_{1}'t}-e^{K_3 t}) + \frac{P_2 M'}{r_{2}'-K_3}(e^{r_{2}'t}-e^{K_3 t}).
\eea
The various terms in the above Eq.~(\ref{un-sol1}) can be made explicit in the following manner:
\bea
P_1&=&\frac{-\kappa L \sqrt{f_{k_l}}}{A+r_1+i(\omega_{m}-\omega_{k_l})}\\
P_2&=&\frac{-\kappa M \sqrt{f_{k_l}}}{A+r_2+i(\omega_{m}-\omega_{k_l})},
\eea
while, $K_1=r_1+i(\omega_{k_l'}-\omega_{k_l})$, $K_2=i(\omega_{k_l'}-\omega_{m})-A$ and $K_3=r_2+i(\omega_{k_l'}-\omega_{k_l})$.
As in the case of the occupied level, the dynamics of the unoccupied level is seen to arise from sums of exponential functions, several of which are 
oscillatory. At time $t=0$, this amplitude vanishes correctly but as time progresses its dynamics is driven again by the three parameters seen 
above in the dynamics of the occupied state. Oscillations at an additional frequency are observed commensurate with the energy difference between the occupied and the unoccupied level
in the left electrode. 

The probability amplitude for the bridge level can be obtained by solving Equation(5) using the Poisson summation, and the initial condition 
of $C_{k_l}(t=0)=\sqrt{f_{k_l}}$. The resultant differential equation has a simple solution in terms of the $C_{k_l}(t)$, namely
\be
C_m(t)=-\frac{i v}{\hbar}\sum_{k_l}\int_0^t dt'\, e^{-A(t-t')}exp[i(\omega_m-\omega_{k_l})t']\,C_{k_l}(t').
\ee
We now solve for $C_m(t)$ using Equations~(\ref{occ-sol1}) \& (\ref{un-sol1}). An important result is that 
we are able to factor out $\sqrt{f_{k_l}}$ from the expressions for $C_{k_l}(t)$, $C_{k_l'}(t)$ and as a consequence also from 
the above expression for $C_m(t)$. Although we have done this only from the solutions to the first recursion it can be 
shown that higher recursions are additive and we can thus always factor out $\sqrt{f_{k_l}}$ from expressions that 
include recursions to infinite order.

Because the microscopic current is defined as the rate of change of the electronic population in the LQC for a given initial condition, 
it can be expressed using the Heisenberg equations of motion as, 
\be
\label{H-I}
\langle \dot{n}_L^{k^0_{l}}\rangle =\frac{2}{\hbar}Im\left\{v\sum_{k_l}C^*_{k_l}(t)C_m(t) e^{i(\omega_m-\omega_{k_l})t}\right\}.
\ee
It is readily shown that the initial occupancy can be factored out from the  $C^*_{k_l}(t)C_m(t)$ product to give Equation(7) in the main 
text, which pertains to a level in the left electrode being occupied initially. Thus 
\be
\label{SklFkl}
\langle \dot{n}_L^{k^0_{l}}\rangle =S_{k^0_{l}}(t)f_{k^0_{l}}.
\ee
where $S_{k^0_{l}}$ has the significance of a reflectance function, since it determines the 
population that remains in the LQC as a result of being reflected back from the bridge 
level. The reflectance function therefore corresponds to an outgoing current from the perspective of the LQC.  Even at the first level of recursion, the  reflectance function can be understood  as a sum of exponential functions with complex arguments. Including higher and higher levels of 
recursions is essentially summing over an infinite number of exponential functions to obtain its most precise analytical form. 

In a similar fashion we obtain the rate of change of population when the initially populated level is in the RQC. We again start with the integro-differential equation for the probability amplitude of a level in the left electrode $\tilde{C}_{k_l}(t)$ when a level in the right electrode is initially populated, i.e., $C_{k_r}(t=0)=\sqrt{f_{k_r}}$,
\bea
\label{full-ID-R}
\dot{\tilde{C}}_{k_l}(t)&=&-\frac{{|v|}^2}{\hbar^2}\left(\int_0^t dt'\,exp[-(A+iB)(t-t')]\,\tilde{C}_{k_l}(t')\right.\nonumber \\
&&+\left.\sum_{k_l'\neq k_l} e^{i\omega_{k_l}t}\int_0^t dt'\,exp[-(A+i\omega_m)(t-t')-i\omega_{k_l'}t']\tilde{C}_{k_l'}(t')\right.\nonumber \\
&&+\left.\sqrt{f_{k_r}}e^{i\omega_{k_l}t}\int_0^t dt'\,exp[-(A+i\omega_m)(t-t')]e^{-i\omega_{k_r}t'} \right).
\eea
Ignoring all levels in the left electrode that are present in the second integral on the right hand side of Equation~(\ref{full-ID-R}) (all  $\tilde{C}_{k_l'}(t')$), enables one to solve for the above as the first recursive solution. 
\bea
\label{simp-ID-R}
\dot{\tilde{C}}_{k_l}(t)&\approx& -\frac{{|v|}^2}{\hbar^2}\left[\int_0^t dt'\,exp[-(A+iB)(t-t')]\,C_{k_l}(t')\right.\nonumber \\
&&+\left.\sqrt{f_{k_r}}\left(\frac{e^{i\omega_{k_lk_r}}t-e^{-(A+i\omega_{mk_l})t}}{A+i\omega_{mk_r}}\right)\right],
\eea
where $\omega_{ab}=\omega_a-\omega_b$. 
As the right electrode population flows into the empty levels of the left electrode, we neglect the flow into all other levels (all $\tilde{C}_{k_l'}(t')$) except the one ($\tilde{C}_{k_l}(t')$) at the level of the first recursive solution.
The solution to equation~(\ref{simp-ID-R}) is
\bea
\label{re-sol}
\tilde{C}^{(1)}_{k_l}&=&-\frac{\kappa\sqrt{f_{k_r}}}{A+i\omega_{mk_r}}\left[\frac{L}{r_1-i\omega_{k_lk_r}}(e^{r_1t}-e^{i\omega_{k_lk_r}t}) 
-\frac{L}{r_1+A+i\omega_{mk_l}}(e^{r_1t}-e^{-(A+i\omega_{mk_l})t})\right.\nonumber \\
 &&+\left.\frac{M}{r_2-i\omega_{k_lk_r}}(e^{r_2t}-e^{i\omega_{k_lk_r}t}) 
-\frac{M}{r_2+A+i\omega_{mk_l}}(e^{r_2t}-e^{-(A+i\omega_{mk_l})t}\right].
\eea
The expression vanishes, as it should, at $t=0$ and and allows a factor $\sqrt{f_{k_r}}$ to be eliminated. 
We obtain an expression for the amplitude of the bridge level, when a level in the right electrode is initially populated, in terms of the probability amplitude of the left electrode levels as,
\be
\label{re-M}
C_m(t)=-\frac{i v}{\hbar}\sum_{k_l}\int_0^t dt'\, e^{-A(t-t')}\left(\tilde{C}_{k_l}(t')\,exp[i(\omega_m-\omega_{k_l})t'] +\sqrt{f_{k_r}}e^{i\omega_{mk_r}t}\right).
\ee.
Employing Equations~(\ref{re-sol})  and (\ref{re-M})  in Equation~(\ref{H-I}) we arrive at the solution in the text, 
\be
\label{SkrFkr}
\langle \dot{n}_L^{k^o_{r}}\rangle =S_{k^0_{r}}(t)f_{k^o_{r}}.
\ee
where $S_{k^0_{r}}(t)$ has the significance of a transmission function.


%
%

%

\bibliography{DynamicalQT.bib}

\end{document}



\title{Supporting Information: Emergence of Landauer Transport from Quantum Dynamics: A Model Hamiltonian Approach} 



\author{Partha Pratim Pal \and S. Ramakrishna \and Tamar Seideman}
\date{%
        Department of Chemistry, Northwestern University, Evanston, IL 60608, USA\\[2ex]
        \today
}





\maketitle 

\begin{figure}[h!]
 \includegraphics[trim={0 8.0cm 0 0},clip, width=0.85\textwidth]{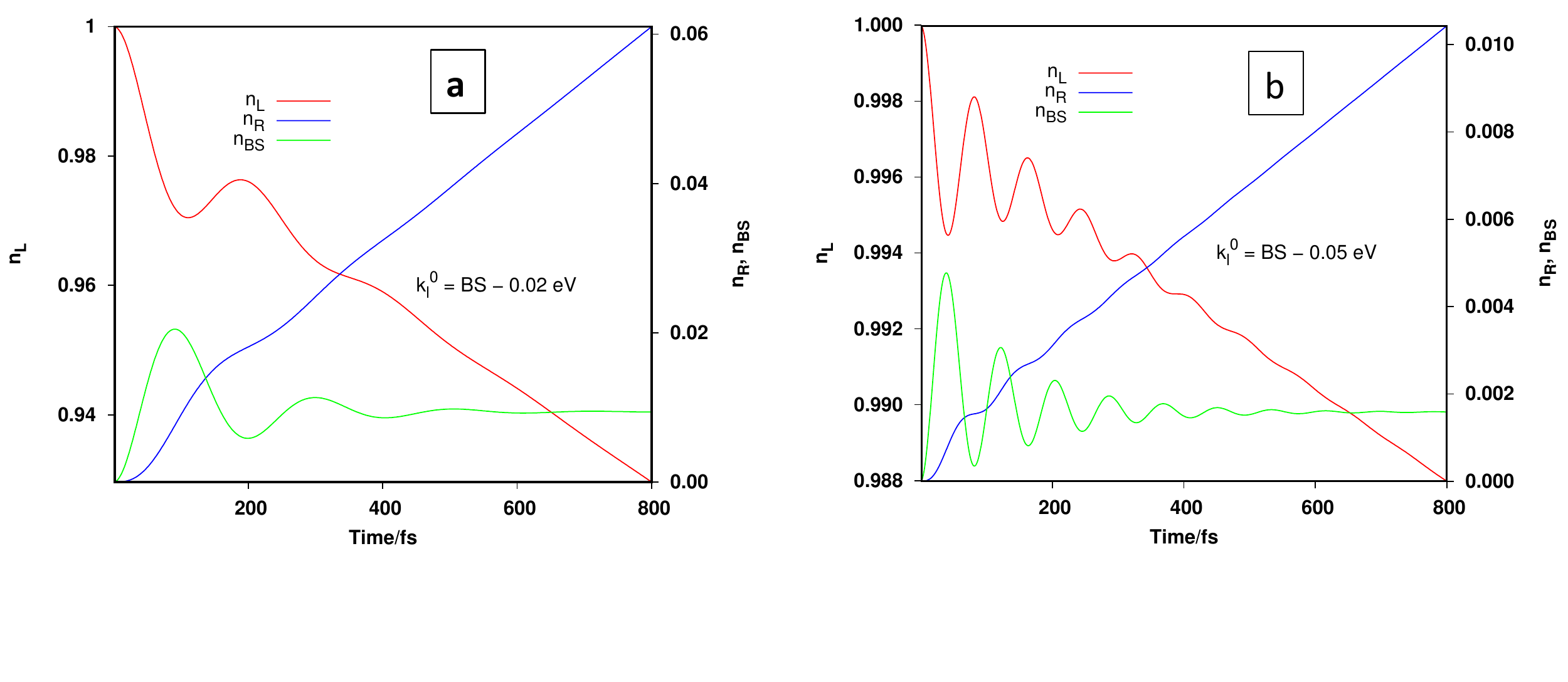}
 \caption{Population dynamics involving a single bridge level connected to two quasi-continua with energy levels from 0 to 1.2 eV. At t = 0, the LQC level at $\epsilon= 0.6$ eV is populated. Populations in the different sub-systems are mentioned in the legends. (a) BS is at 0.62 eV. (b) BS is at 0.65 eV.}
 \label{figSI:single}
\end{figure}

We briefly summarize the single particle dynamics within the framework introduced in IIA 
of the main article. The QC energy levels range from 0 to 1.2 eV and a coupling strength of 0.002 eV is used in the calculations. 
The LQC level at $\epsilon = 0.6$~eV is populated at t = 0 for both calculations but the position of BS is varied as indicated in the caption of Figure SI-\ref{figSI:single}. 
We see that the frequency of oscillations in the population of all the sub-systems is dependent on the energy difference between the initially populated level and the BS.
As expected, the larger the difference the more rapid the fluctuation (see also equations 3, 4, and 5 in the main article).
The oscillations are damped as population in the BS saturates to a constant value and the system transitions into a dynamic equilibrium.

\begin{figure}[h!]
 \includegraphics[trim={0 0.0cm 0 0},clip, width=0.85\textwidth]{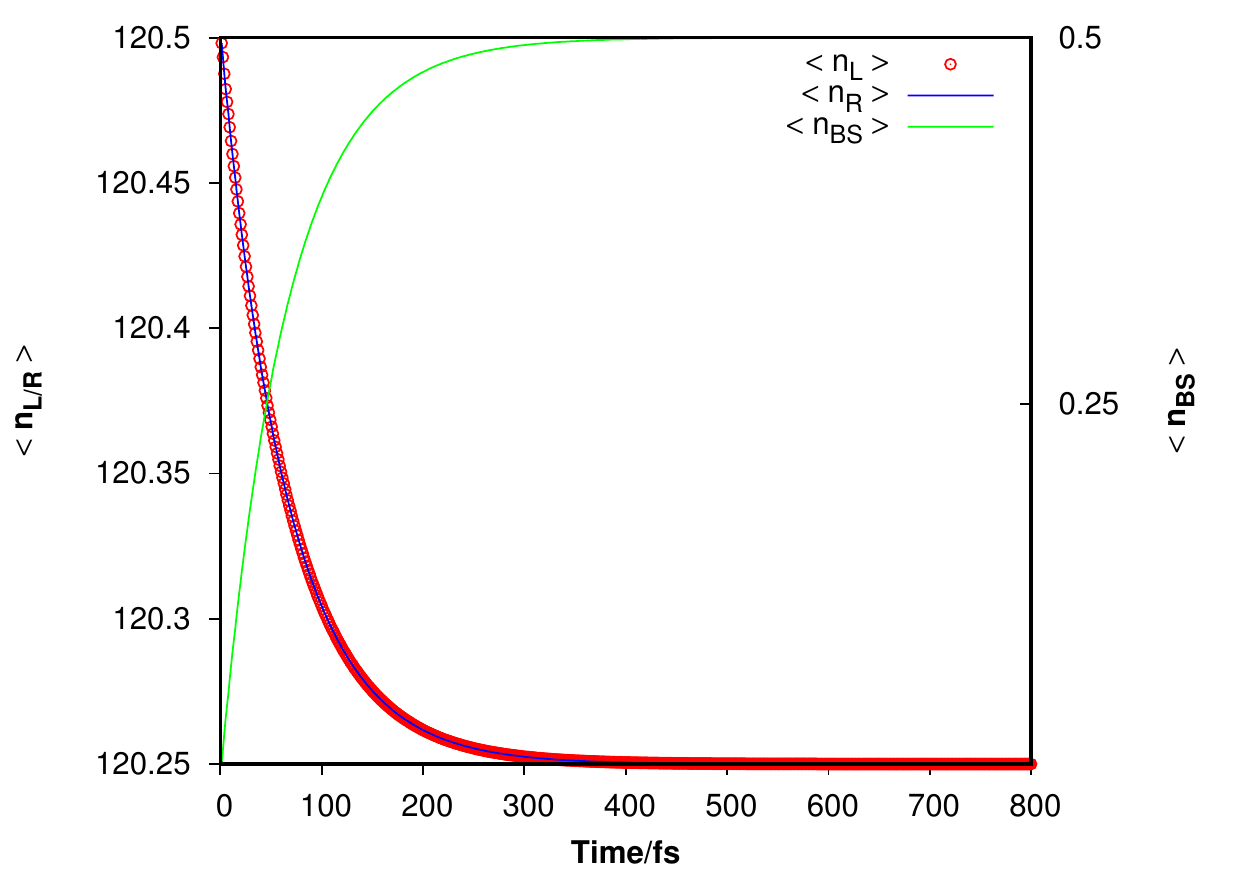}
 \caption{Population dynamics involving a single bridge level connected to two quasi-continua where all the levels in the quasi-continua are populated one by one at t=0. Populations in the different sub-systems are mentioned in the legends.}
 \label{figSI1:pop_dyn}
\end{figure}

\begin{figure}[h!]
 \includegraphics[trim={0 0.0cm 0 0},clip, width=0.85\textwidth]{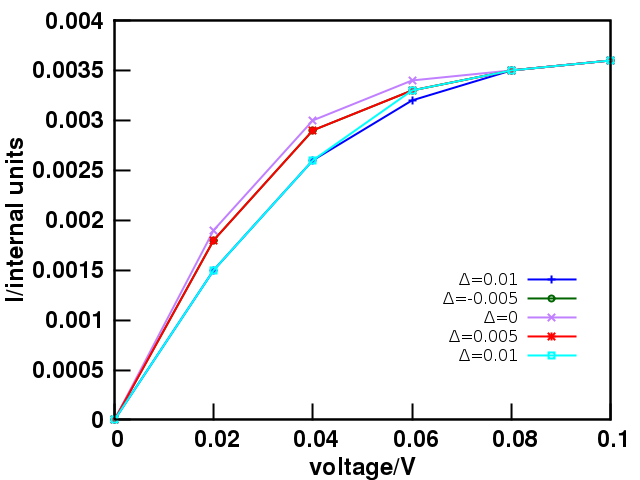}
 \caption{Current-voltage characteristics for different positions of BS. `$\Delta$' denotes the shift of BS from the equilibrium $E_f$.}
 \label{figSI:pos_BS}
\end{figure}

\begin{figure}[h!]
 \includegraphics[trim={0 0.0cm 0 0},clip, width=0.85\textwidth]{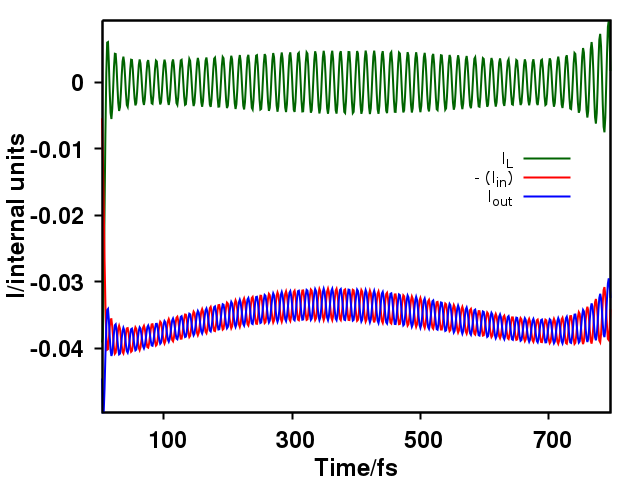}
 \caption{I$_{in}$ and I$_{out}$ at LQC for a strong coupling, illustrating  oscillations in both $I_{in}$ and $I_{out}$. 
	}
 \label{figSI:It_ec001}
\end{figure}

\begin{figure}[h!]
 \includegraphics[trim={0 0.0cm 0 0},clip, width=0.85\textwidth]{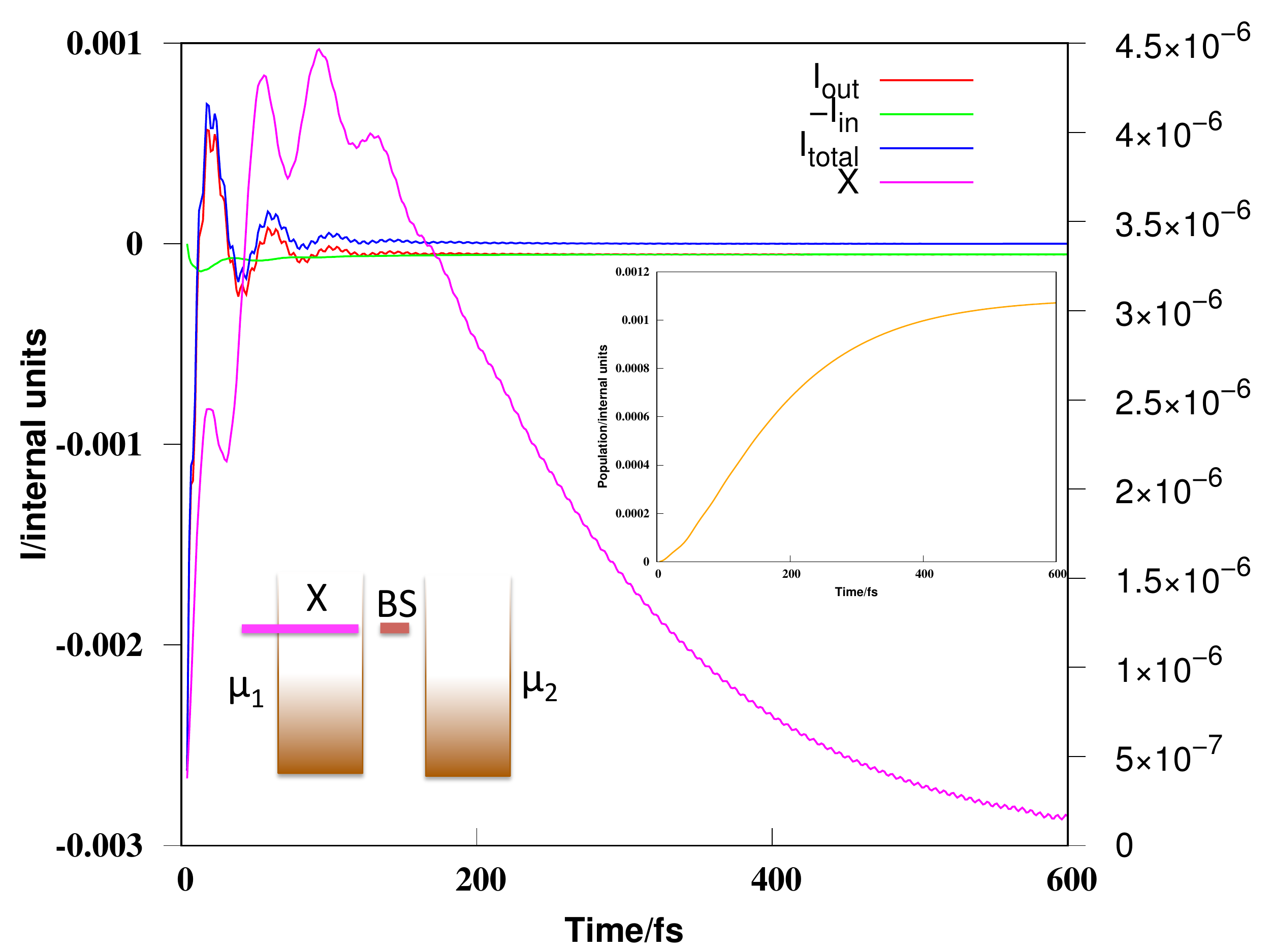}
 \caption{I$_{in}$, I$_{out}$, and $I_{L}$ when BS = CP + 0.1 eV. A magenta curve denotes the rate of change of population (plotted against the right ordinate) of the level marked as `X' in the schematic. `X' is resonant with BS. Inset: Population at the `X' level as a function of time.  
}
 \label{figSI:OffR_analysis}
\end{figure}

\begin{figure}[h!]
 \includegraphics[width=0.85\textwidth]{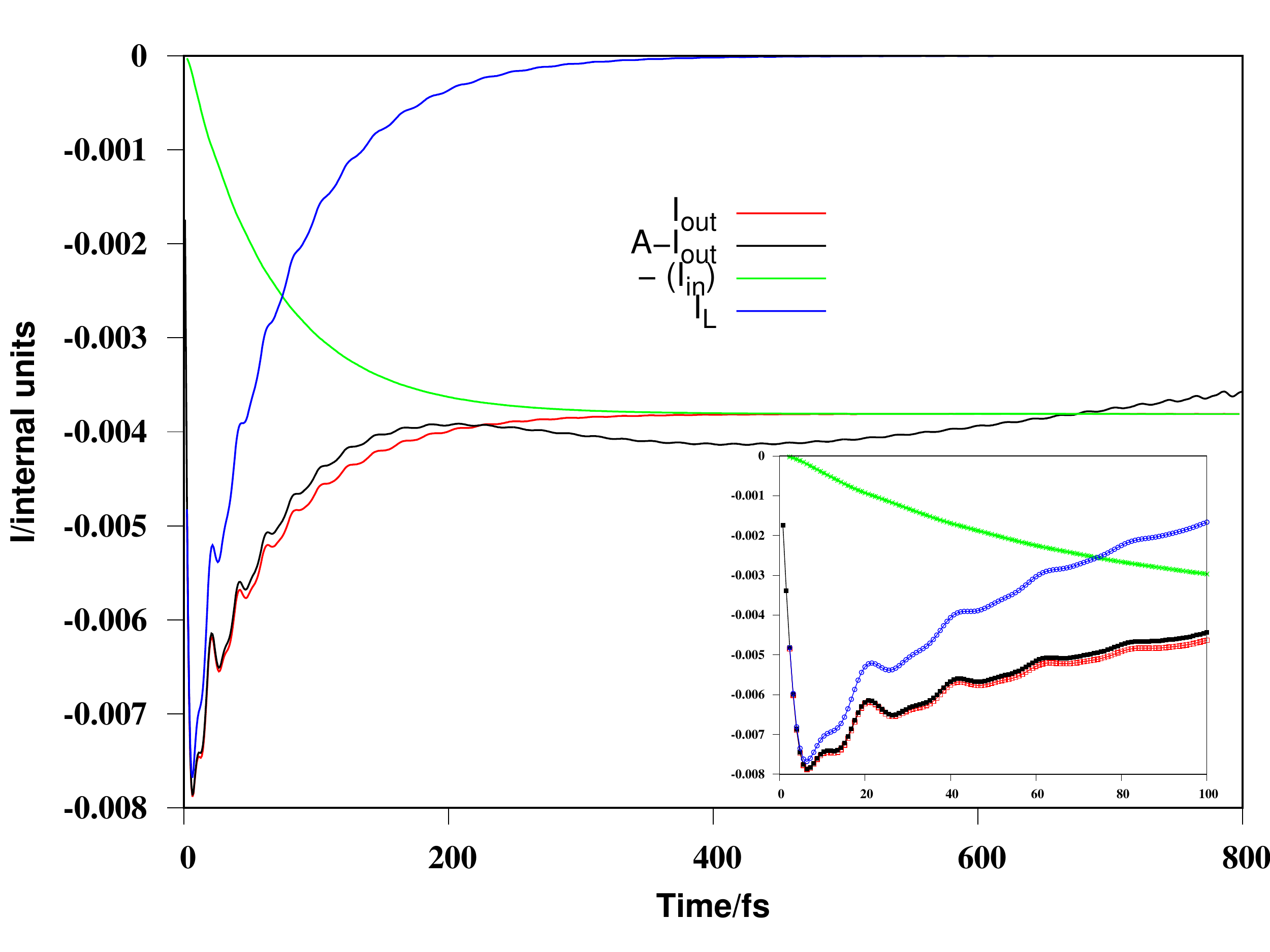}
 \caption{Population dynamics at the LQC and its components $I_{out}$ and -$I_{in}$. The bridge level is 0.4 eV below the Fermi level of the quasi-continua. The definition of the components are provided in the text. Inset: Population dynamics at very early times.}
 \label{figSI:mp-noff}
\end{figure}

\begin{figure}[h!]
 \includegraphics[trim={6.0cm 0.0cm 6.0cm 0},clip, width=0.85\textwidth]{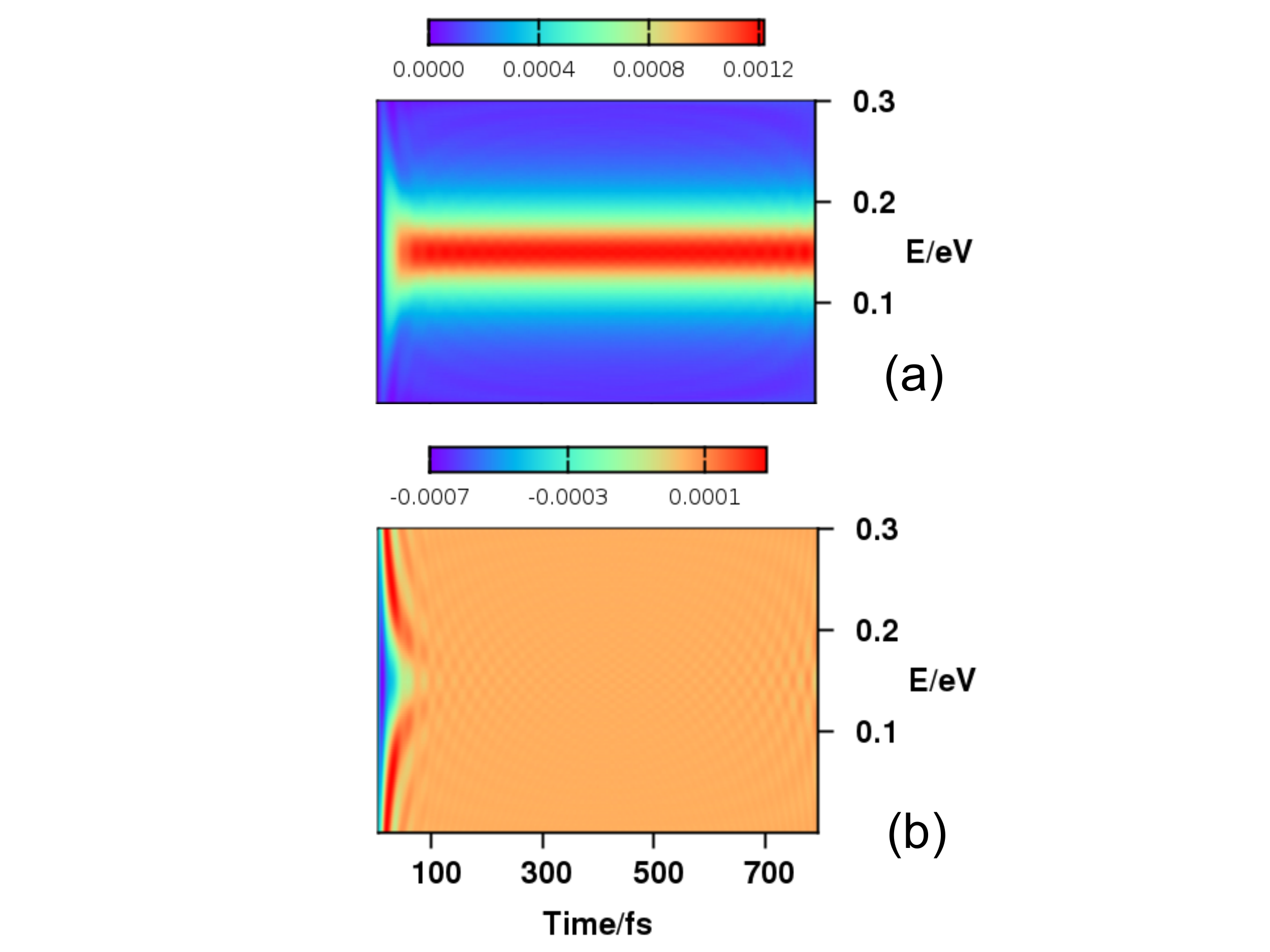}
 \caption{ (a) The transmission function ($S_{k^0_r}$) as a function of time and energy. (b) The difference between $S_{k^0_l}$ and $S_{k^0_r}$ as a function of time and energy.}
 \label{figSI:trans}
\end{figure}

\begin{figure}[h!]
 \includegraphics[trim={0 0.0cm 0 0},clip, width=0.85\textwidth]{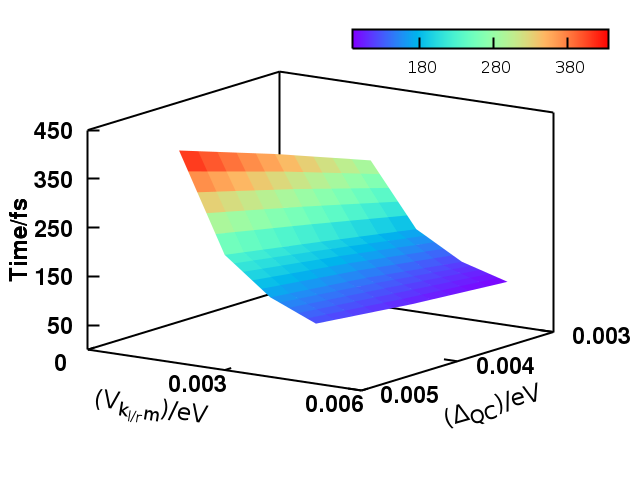}
 \caption{Onset of the Landauer regime for a range of symmetric coupling  strengths between the bridge level and the quasi-continua. It is calculated from the approximate Landauer type expression given by equation 10b.
 The bridge level is on-resonance with the Fermi level and the bias voltage is taken to be 0.02 V.}
 \label{figSI:onset_L}
\end{figure}


%
%

%

